\documentclass[epjST]{svjour}

\usepackage[utf8]{inputenc}
\DeclareUnicodeCharacter{2212}{-}
\DeclareUnicodeCharacter{2032}{'}
\usepackage{amsmath}
\usepackage{amssymb}
\usepackage{parskip}

\usepackage[
natbib=true,
style=numeric-comp,
sorting=none]{biblatex}
\usepackage{graphicx}
\usepackage{dcolumn}
\usepackage{bm}
\usepackage{hyperref}
\hypersetup{
  colorlinks   = true, 
  urlcolor     = blue, 
  linkcolor    = blue, 
  citecolor   = red 
}

\usepackage{slashed}
\usepackage{braket}

\def\be{\begin{equation}}
\def\ee{\end{equation}}
\def\bea{\begin{eqnarray}}
\def\eea{\end{eqnarray}}
\def\bear{\begin{array}}
\def\ear{\end{array}}
\def\bfig{\begin{figure}}
\def\efig{\end{figure}}
\def\bcen{\begin{center}}
\def\ecen{\end{center}}
\def\bi{\begin{itemize}}
\def\ei{\end{itemize}}
\def\raw{\rightarrow}

\def\bpi{\bm\pi}

\begin{document}
\title{Neutrino Interactions with Matter and the MiniBooNE anomaly}
\author{Luis Alvarez-Ruso and Eduardo Saul-Sala}
\institute{Instituto de F\'isica Corpuscular (IFIC) and Departamento de F\'\i sica Te\'orica, \\ Consejo Superior de Investigaciones Cient\'{i}ficas (CSIC) and Universidad de Valencia (UV)\\ E-46980 Paterna, Valencia, Spain}
\abstract{
The excess of electron-like events measured by MiniBooNE challenges our understanding of neutrinos and their interactions. We review the status of this open problem and ongoing efforts to resolve it. After introducing the experiment and its results, we consider the main experimental backgrounds and the related physics of neutrino interactions with matter such as quasielastic-like scattering and weak pion production on nucleons and nuclei. Special attention is paid to single photon emission in neutral current interactions and, in particular, its coherent channel. The difficulties to reconcile the MiniBooNE anomaly with global oscillation analysis is then highlighted. We finally outline some of the proposed  solutions of the puzzle involving unconventional neutrino-interaction mechanisms.     
} 
\maketitle

\section{The MiniBooNE short baseline anomaly}
\label{sec:theano}

Neutrinos and antineutrinos are emitted in weak processes as flavor eigenstates. Once these are linear combinations of mass eigenstates, (anti)neutrinos change flavor with time because the phases of mass eigenstates evolve differently. Oscillation experiments detect charged leptons originated in (anti)neutrino charged-current (CC) interactions with a target. As an oscillation signature, appearance measurements search for charged leptons with flavors that differ from those of the originally produced neutrinos. MiniBooNE and the earlier LSND experiments belong to this category.     

\paragraph{Precedent: LSND.}

The Liquid Scintillator Neutrino Detector (LSND) was illuminated by a beam of electron and muon (anti)neutrinos produced at Los Alamos National Laboratory from pion and muon decay at rest: $\pi^+ \raw \mu^+ \, \nu_\mu$ followed by $\mu^+ \raw e^+ \, \nu_e \, \bar\nu_\mu$. Electron antineutrinos were revealed by inverse beta decay $\bar\nu_e \, p \raw e^+ \, n$. During its operation time between 1993 and 1998, the experiment found a signal excess of $87.9 \pm 22.4 \pm 6.0$ $\bar\nu_e$ events over the expected small intrinsic background~\cite{Aguilar:2001ty}. In a simple model with two mass eigenstates, the oscillation probability is
\be
P = \sin^2{2 \theta} \sin^2{\left( 1.27 \Delta m^2 \frac{L}{E_\nu} \right)}
\ee
with $\Delta m^2 = |m_1^2 - m_2^2|$ in units of eV$^2$, $L$ the distance traveled by the neutrino in meters and $E_\nu$ its energy in MeV. 
Attributing the excess to $\bar\nu_\mu \raw \bar\nu_e$ short ($\sim 30$~m) baseline oscillations, an allowed region in the $(\sin^2{2 \theta}, \Delta m^2 )$ plane  is obtained with a best-fit $\Delta m^2 = 1.2$~eV$^2$ (see for instance Fig.~26 of Ref.~\cite{Aguilar:2001ty}). Such a mass splitting, much larger than those obtained in the three-flavor paradigm established from solar, atmospheric, reactor and accelerator experiments~\cite{ParticleDataGroup:2020ssz}, can be in principle accommodated by the introduction of sterile neutrinos that mix with the Standard Model (SM) flavors but do not couple directly to the weak bosons.   

\paragraph{The MiniBooNE experiment.}
The MiniBooNE experiment~\cite{MiniBooNE:2008paa} at Fermilab, was designed to detect electron (anti)neutrinos in a muon (anti)neutrino beam, with an average $L/E_\nu \sim 1$~m/MeV, similar to LSND, in order to test its earlier result.

\begin{figure}[htbp]
    \centering
        \includegraphics[width=0.45\textwidth]{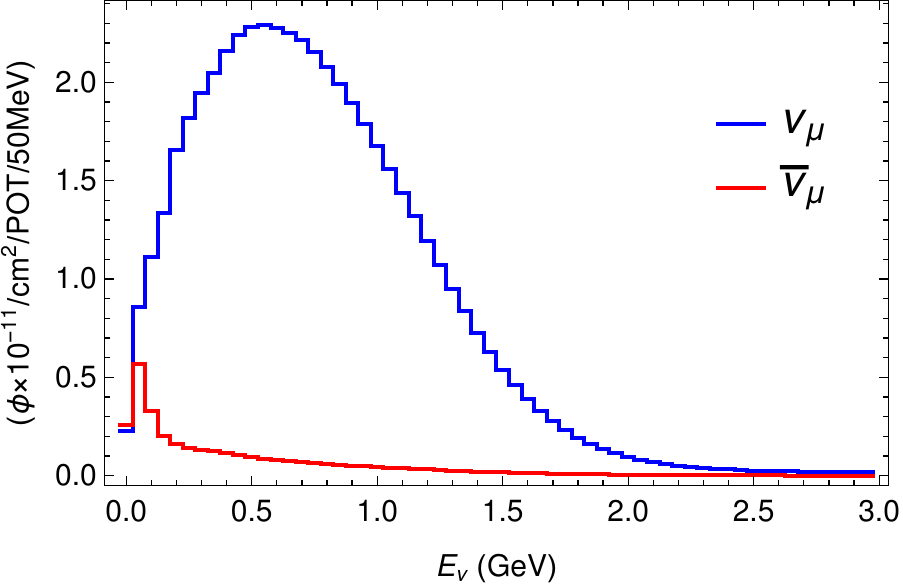}
    ~$ \hspace{0.3cm} $~
        \includegraphics[width=0.45\textwidth]{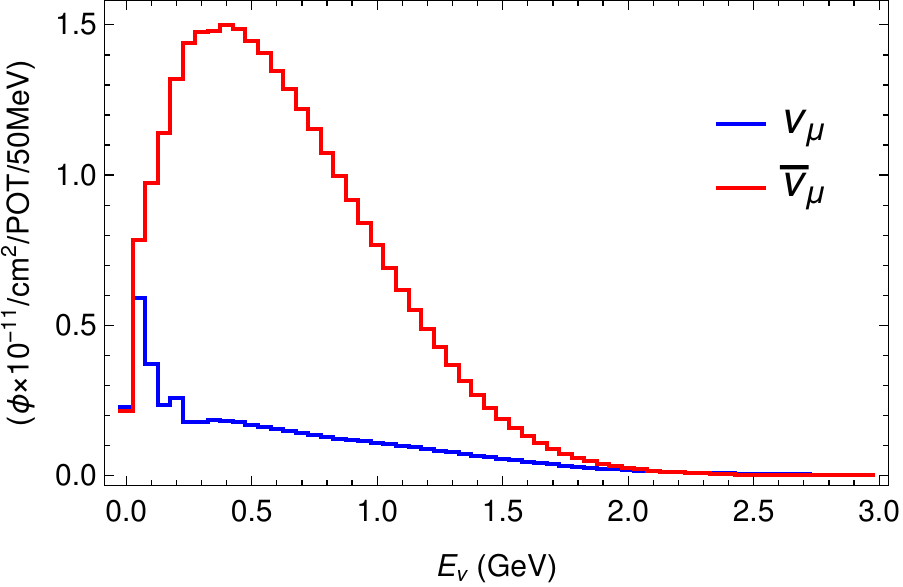}
    \caption{Leading components of the neutrino flux at MiniBooNE in neutrino (left) and antineutrino (right) modes~\cite{AguilarArevalo:2008yp}.}
    \label{ch:H_nu:sec:MiniBooNE:fig:MiniBooNE_flux}
\end{figure}

The MiniBooNE detector is a 12.2 m diameter spherical tank filled with 818 tonnes of mineral oil, CH$_2$, with 1280 photomultiplier tubes to collect Cherenkov light produced by charged particles emitted in the interaction processes. The neutrino flux directed to MiniBooNE is produced by meson (mostly pion) decay in flight at the Fermilab Booster neutrino beamline, with a baseline of  541 m. The 20 times larger baseline compared to LSND entails a proportionally larger $E_\nu \sim 500$~MeV to keep $L/E_\nu \sim 1$~m/MeV. 
The proton beam is directed to a beryllium target, where the secondary meson beam is produced. Using magnetic fields, one of the components of the meson beam can be selected to obtain a beam of predominantly neutrinos or antineutrinos.
For example, by keeping the $\pi^+$ muonic neutrinos are favored by their decay, $\pi^+ \rightarrow \mu^+ + \nu_\mu$. 
The largest $\nu_\mu$ and   $\bar{\nu}_\mu$ components of the fluxes at the MiniBooNE detector in both neutrino and antineutrino modes  are shown in Fig.~\ref{ch:H_nu:sec:MiniBooNE:fig:MiniBooNE_flux}. Electron (anti)neutrino components are orders of magnitude smaller~\cite{AguilarArevalo:2008yp}. It can be seen that the $\nu_\mu$ contamination in antineutrino mode is relatively larger than the $\bar\nu_\mu$ one in neutrino mode. 

\begin{figure}[ht]
    \centering
    \includegraphics[width=0.65\textwidth]{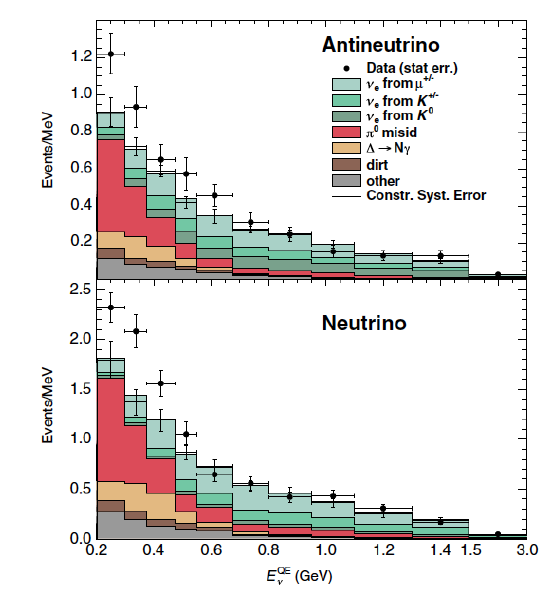}
    \caption{Results of the MiniBooNE experiment in both antineutrino (upper panel) and neutrino (lower panel) modes~\cite{Aguilar-Arevalo:2013pmq}.
    The distribution of electron-like events (oscillation candidates) as a function of $E_\nu^\mathrm{QE}$ is shown together with background estimates.}
    \label{fig:excess_miniboone}
\end{figure}

The experiment searched for electron-like charged-current quasielastic (CCQE) events originated in $\nu_e \, n \raw e^- \, p$ ($\bar\nu_e \, p \raw e^+ \, n$) interactions in neutrino (antineutrino) mode. Data collected between 2002 and 2012 for $6.46 \times 10^{20}$ protons on target (POT) in neutrino mode and $11.27 \times 10^{20}$~POT in antineutrino mode showed an excess of events over the predicted background in both cases~\cite{Aguilar-Arevalo:2013pmq}. The distribution of these events are shown in Fig.~\ref{fig:excess_miniboone} 
as a function of $E_\nu^\mathrm{QE}$, defined as the energy of the incoming neutrino reconstructed from the energy $E_e$ and scattering angle $\theta_e$ of the final $e^\pm$ assuming that the interaction took place on a single non-interacting nucleon bound with a constant binding energy
\begin{equation}
\label{eq:Enurec}
E_\nu^\mathrm{QE} = \frac{2 M'_{n(p)} E_e - M'^2_{n(p)} + M_{p(n)}^2 -m_e^2}{2 (M'_{n(p)} - E_e +\sqrt{E_e^2 - m_e^2} \cos{\theta_e})}\,,    
\end{equation}
where $M'_{n(p)} = M_{n(p)} - E_B$ with $E_B = 34$~MeV. 
The excess is concentrated at $200 < E_\nu^{\mathrm{QE}} < 475$~MeV. The lower $E_\nu^{\mathrm{QE}}$ limit is dictated by the ability to reconstruct reliably $\nu_\mu$ Cherenkov rings with visible energies greater than 140~MeV.  
From 2012 to 2019, data has been further collected in neutrino mode, reaching $18.75 \times 10^{20}$~POT~\cite{MiniBooNE:2020pnu}, confirming the original results (compare Fig.~9 of Ref.~\cite{MiniBooNE:2020pnu} to the lower plot in Fig~\ref{fig:excess_miniboone} from Ref.~\cite{Aguilar-Arevalo:2013pmq}). 

\begin{figure}[ht]
    \centering
    \includegraphics[width=0.65\textwidth]{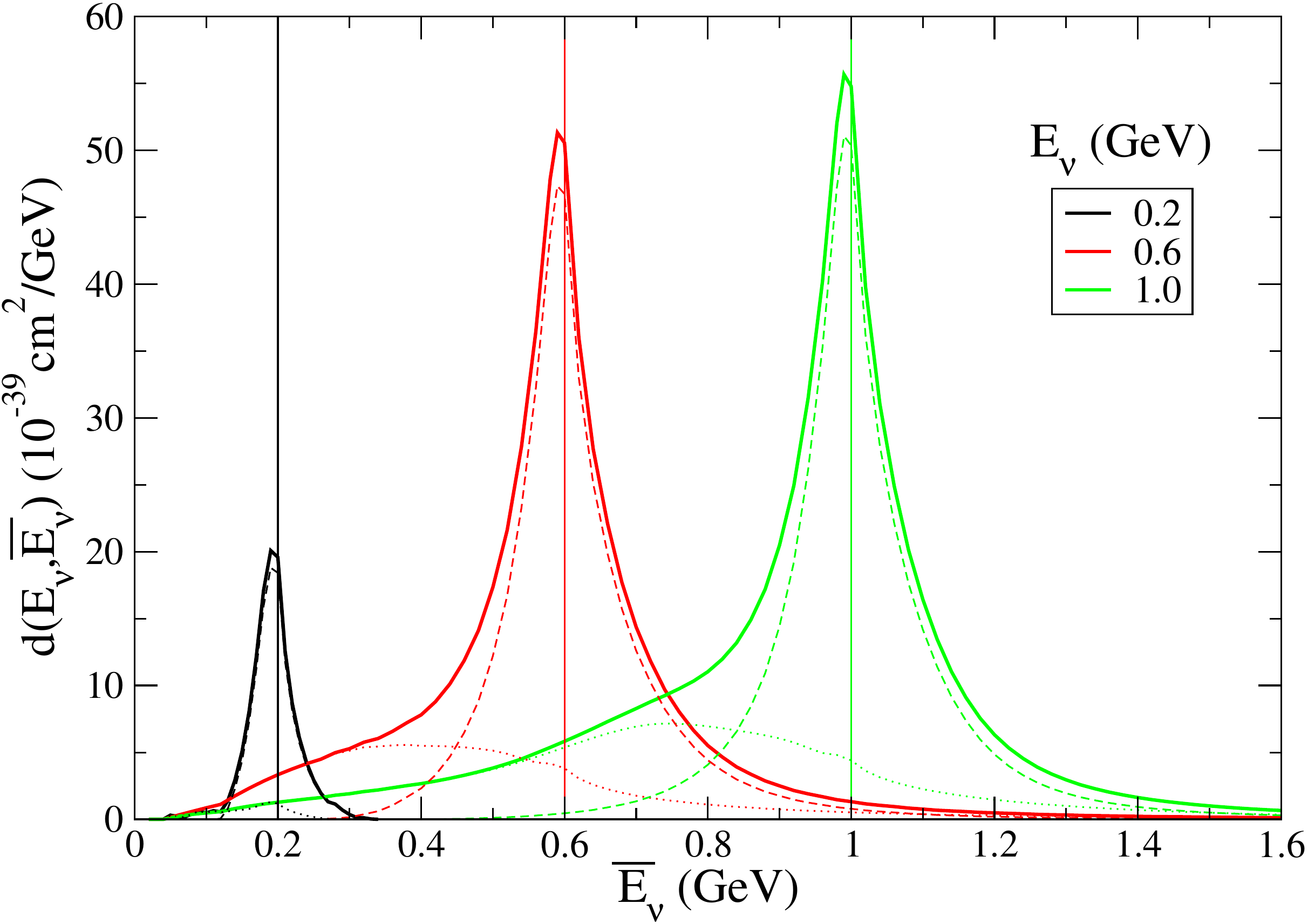}
    \caption{Spreading of $E_\nu^{\mathrm{QE}}$ (denoted $\bar{E}_\nu$) in $\nu_e$-$^{12}$C CCQE-like scattering due to nuclear effects, for three fixed true $E_\nu$~\cite{Martini:2012uc}.}
    \label{fig:Martini_Erec}
\end{figure}

An oversimplification of nuclear structure and reaction dynamics underlies in the experimentally adopted definition of $E_\nu^{\mathrm{QE}}$ given above. It is known that nucleons in the nucleus are not at rest but undergo Fermi motion and, when knocked out from the nucleus, propagate in density and momentum-dependent mean field potentials (see for instance the discussion in section 2.5 of Ref.~\cite{Mosel:2019vhx}). Moreover, external probes can interact with nucleon pairs or, at low energy and momentum transfers, collective modes can be excited~\cite{Jachowicz:2019eul}. Several independent calculations have established that a sizable fraction of the CCQE-like dataset collected by MiniBooNE originates in interactions with nucleon pairs~\cite{Martini:2011wp, Nieves:2011yp, Megias:2014qva,Lovato:2020kba}. This introduces a bias in the migration matrix form $E_\nu^{\mathrm{QE}}$ to the true $E_\nu$~\cite{Martini:2012uc, Nieves:2012yz}, as can be appreciated in Fig~\ref{fig:Martini_Erec} taken from Ref.~\cite{Martini:2012uc}. 
Besides Fermi smearing, low energy tails due to scattering off nucleon pairs are present at all but the lowest $E_\nu = 200$~MeV. Additional strength in the tail arises from CC pion production events followed by pion absorption in the nucleus~\cite{Lalakulich:2012hs}. 

While the presence of such a bias should not cast doubts on the existence of the electron-like excess itself, it might influence its explanation. In particular, if interpreted in terms of $\nu_\mu \raw \nu_e$ oscillations, an unaccounted mismatch between the true $E_\nu$, upon which the oscillation probabilities depend, and $E_\nu^{\mathrm{QE}}$ will alter the determination of oscillation parameters. For tests of alternative explanations of the MiniBooNE anomaly not involving neutrino oscillations, a representation of the excess of events in terms of the visible energy and angle between the reconstructed electron and the beam direction (see Figs.~7, 8 and 13 of Ref.~\cite{MiniBooNE:2020pnu}) is better suited.  

\section{Backgrounds at MiniBooNE}
\label{sec:thebackg}

As apparent from Fig.~\ref{fig:excess_miniboone}, the mere existence, size and kinematic distribution of the MiniBooNE excess of events critically relies on the proper determination of electron-like backgrounds which, in the absence of a near detector, are determined using MiniBooNE data. A brief updated description of these backgrounds has been presented in Ref.~\cite{Katori:2020tvv}; here we revisit them from a more theoretical perspective.

\paragraph{Intrinsic electron-(anti)neutrino background.} This background comes from in-flight decays of muons and kaons. The $\nu_e \, (\bar\nu_e)$ component in the flux from muon decays is directly related to the observed $\nu_\mu \, (\bar\nu_\mu)$ events due to the common origin in pion decays. The fraction from kaon decays is constrained by fits to kaon production data and by high-energy data measured at the SciBooNE detector~\cite{SciBooNE:2011sjq}. The number of CCQE-like events at the detector and their kinematic distributions are determined by the experiment's Monte Carlo, tuned to measured $\nu_\mu$ CCQE-like scattering on $^{12}$C. However, the Monte Carlo simulation does not take into account multinucleon-events. This is a potential source of systematic uncertainty as the $\nu_\mu/\nu_e$ cross section ratio for scattering on nucleon pairs is different than for CCQE scattering. 
\begin{figure}[h!]
    \centering
        \includegraphics[width=0.45\textwidth]{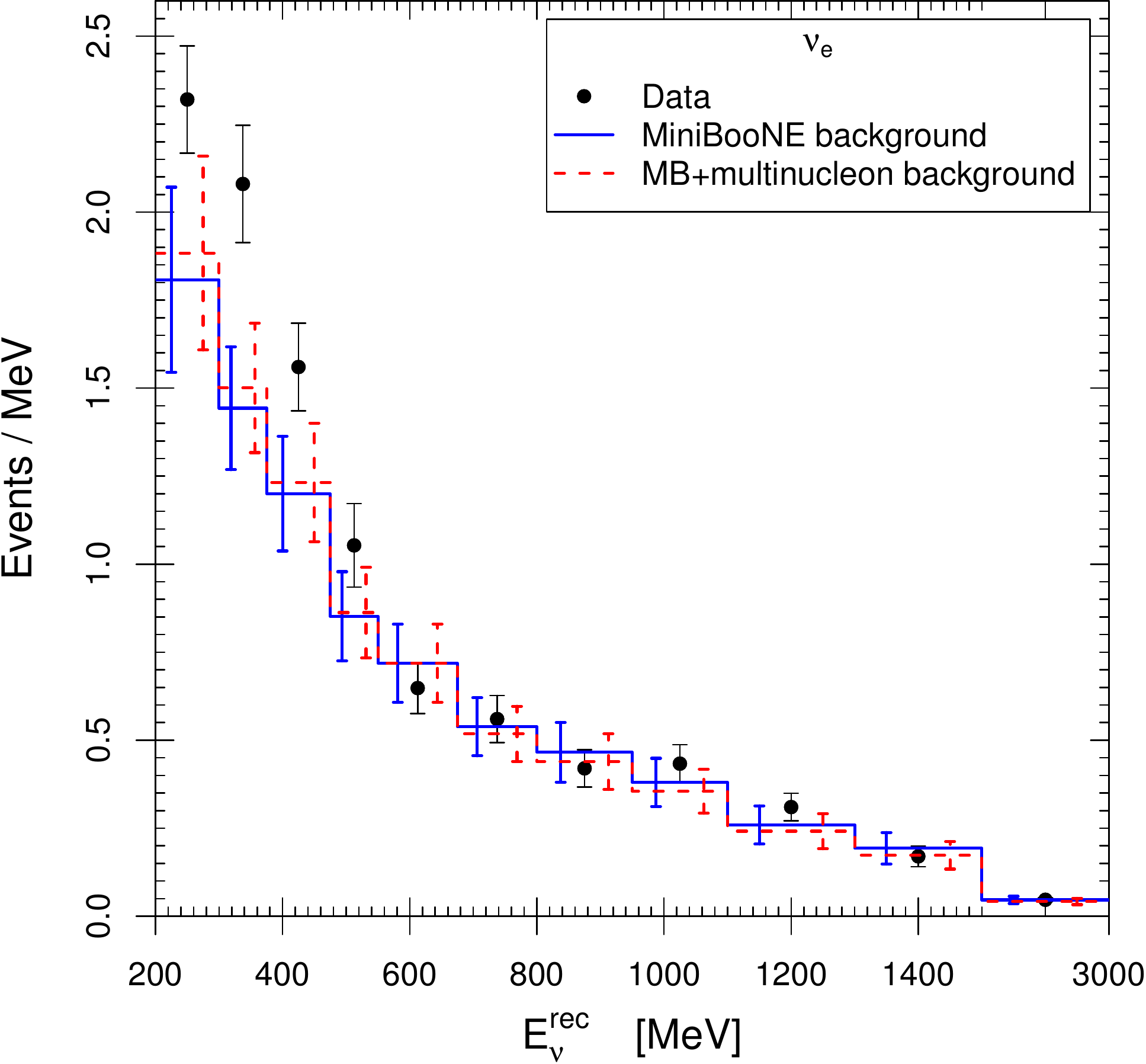}
    ~$ \hspace{0.3cm} $~
        \includegraphics[width=0.45\textwidth]{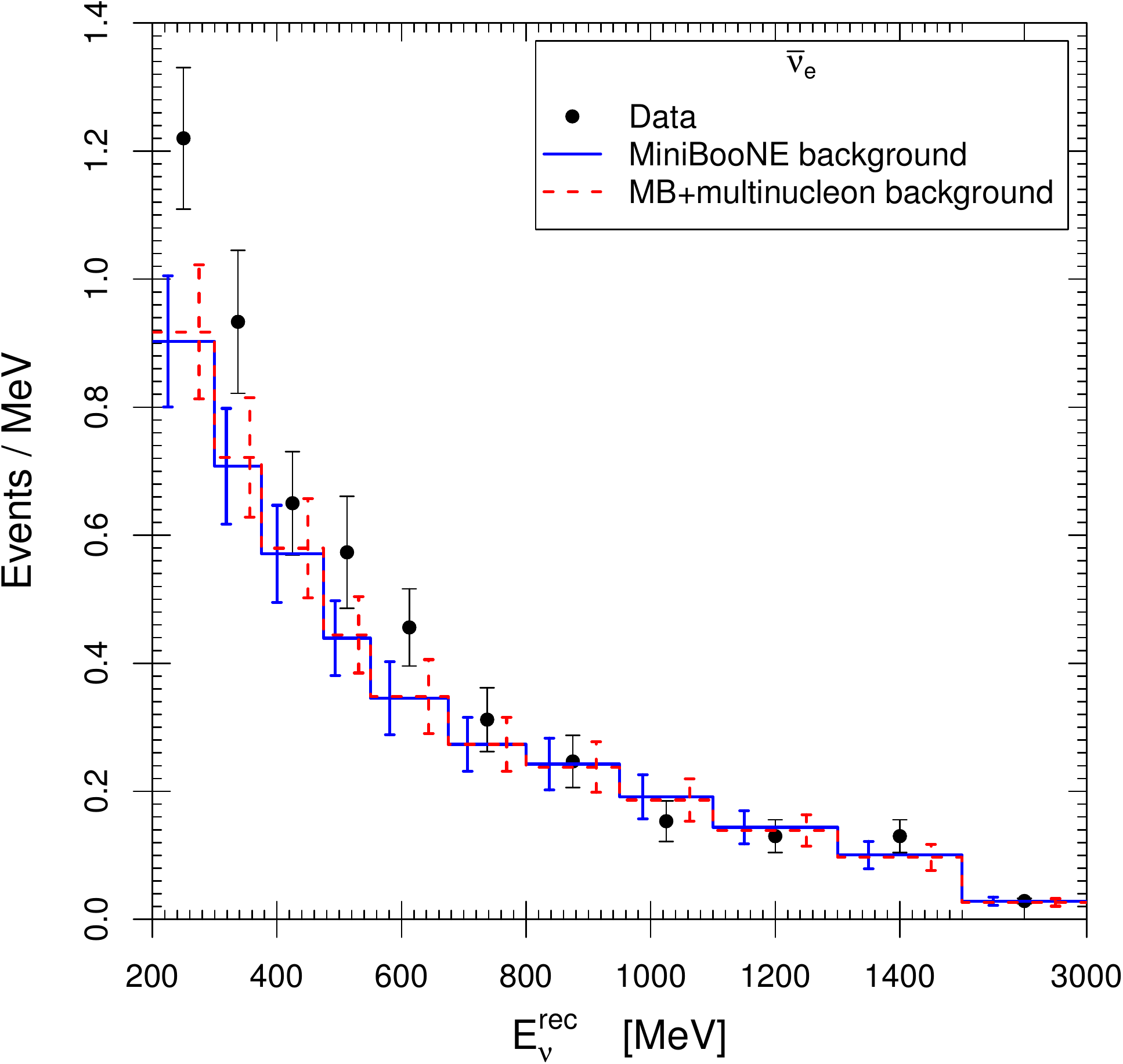}
    \caption{Electron-like background at MiniBooNE taking into account multinucleon contributions~\cite{Ericson:2016yjn} compared to the original estimate.}
    \label{fig:nue_background}
\end{figure}
The impact of this difference on the shape (not in the total number of $\nu_e \, (\bar\nu_e)$, which is normalized to the MiniBooNE prediction) of the predicted intrinsic $\nu_e \, (\bar\nu_e)$ background was studied in Ref.~\cite{Ericson:2016yjn}. After this correction, the obtained background, shown in Fig.~\ref{fig:nue_background}, is consistent with the original MiniBooNE estimate but shows an enhancement in the low energy bins, which is stronger in neutrino mode.

\begin{figure}[h!]
    \centering
    \includegraphics[width=0.65\textwidth]{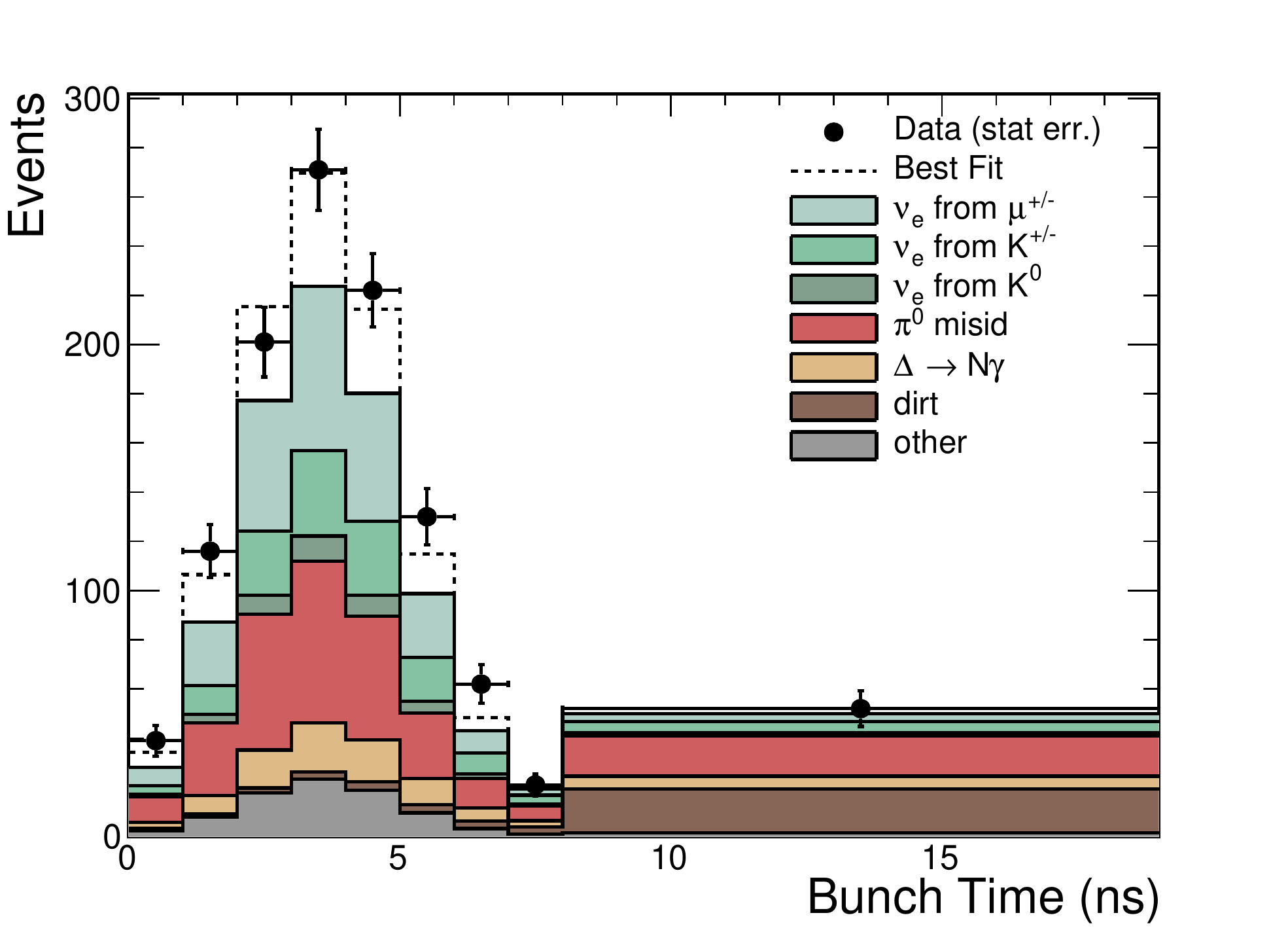}
    \caption{The bunch timing for data events in neutrino mode compared to the expected background~\cite{MiniBooNE:2020pnu}.}
    \label{fig:timebunch}
\end{figure}

\paragraph{External events.} Photons produced outside the detector can give a signal inside without triggering a veto. As Cherenkov detectors cannot distinguish between photons and electrons, such signals are indistinguishable from CCQE-like interactions of electron (anti) neutrinos. These {\it dirt} background events are difficult to estimate as the outside surroundings of the detector are incompletely simulated. For this purpose, MiniBooNE isolates events near  the  edge of  the  detector  and  pointing  towards  the  detector  center. Furthermore, timing information (see Fig.~\ref{fig:timebunch} taken from Ref.~\cite{MiniBooNE:2020pnu}) shows that the event excess peaks in the 8~ns window associated with beam bunch time, as expected from neutrino events in the detector. The {\it dirt} event prediction is tested by the correct description of data taken off-phase with respect to the beam. 

\paragraph{Neutral-current $\bpi^{\bm 0}$ background.} The production of neutral pions by neutral current (NC) interactions of (anti)neutrinos of all flavors with nuclei in the detector material $\nu (\bar\nu) \, A \raw \nu (\bar\nu)  \, X \, \pi^0$ (the hadronic final state $X$ contains any number of nucleons and nuclear fragments but no mesons or photons) is a source of electron-like background events in Cherenkov detectors such as MiniBooNE or Super Kamiokande when one of the photons from the $\pi^0 \raw \gamma \, \gamma$ decay is unresolved. In the case of MiniBooNE this background, represented in red in Fig.~\ref{fig:excess_miniboone}, is the largest at low $E_\nu^{\mathrm{QE}}$, where the excess of events is present. 

A reliable simulation of this background should tackle (at least to some extent) the challenging problem of weak pion production in nuclei. A basic ingredient is a realistic model for neutrino-induced pion production on single nucleons. Although this process is dominated by the excitation of baryon resonances and their subsequent decay into $\pi N$, there are also non-resonant mechanisms which coexist and interfere with resonant ones. At MiniBooNE energies, the largest contribution is mediated by the $\Delta(1232)$ resonance. Isospin considerations imply that the $\Delta(1232)$ is dominant for $\nu_l p \raw l^- \, p \, \pi^+$ and $\bar\nu_l n \raw l^+ \, n \, \pi^-$ but the relative importance of non-resonant and $N^*$-mediated amplitudes is larger for NC reaction channels~\footnote{The $W^+ \, p \raw \Delta^{++} \raw p \, \pi^+$ matrix element is proportional to Clebsch-Gordan coefficient $(1/2 \,1/2 \, 1 \,1 | 3/2 \, 3/2)^2 = 1$ while for $Z^0 \, p \raw \Delta^{+} \raw p \, \pi^0$ one has $(1/2 \,1/2 \, 1 \,0 | 3/2 \, 1/2)^2 = 2/3$.}. While at low energy and momentum transfers, weak pion production can be systematically studied using Chiral Perturbation Theory~\cite{Yao:2019avf}, the kinematic range probed by neutrino interactions at MiniBooNE demands a more phenomenological approach. Phenomenological models for weak meson production rely on symmetries to constrain the parameters with precise and abundant non-neutrino data. Owing to isospin symmetry, the form factors that characterize the vector part of the weak current can be extracted from pion electroproduction data~\cite{Leitner:2008ue}. On the other hand, the partial conservation of the axial current (exact only in the chiral limit but still a good approximation thanks to the lightness of pions) allows to relate the axial current at zero four-momentum transfer squared ($q^2$) to the pion-nucleon scattering amplitude, which is also well known experimentally. Such a connection with pion electroproduction and pion-nucleon scattering data is present in all models but is most extensively exploited by the dynamical model in coupled channels of Ref.~\cite{Nakamura:2015rta}. What remains unconstrained by non-neutrino data is the $q^2$ dependence of the axial current form factors, on which only limited information can be extracted from bubble-chamber data on pion production induced by neutrinos on deuterium, taken at Argonne and Brookhaven National Laboratories (ANL and BNL)~\cite{Hernandez:2010bx}.

When pions are produced on  nuclear  targets as in all modern neutrino experiments, including MiniBooNE,  the  presence  of  the  nuclear medium poses additional challenges for the reaction modeling. Given the prevalent role of the $\Delta(1232)$ excitation in pion production, it is not surprising that the in-medium modification of the $\Delta$ propagator is very important. The main effect is the increase of the $\Delta(1232)$  width (broadening) by many body processes: $\Delta \,N \raw N \, N$, $\Delta \,N \raw N \, N \, \pi$, $\Delta \,N \, N \raw N \, N \, N$. In their way out of the nucleus, pions undergo final state interactions (FSI). They can be absorbed, change their energy, angle and charge. In particular, in NC interactions, there is a shift of strength from the largest $\pi^0$ production channel to the $\pi^\pm$ ones via charge-exchange FSI: $\pi^0 \, p \raw \pi^+ \, n$ and $\pi^0 \, n \raw \pi^- \, p$~\cite{Leitner:2006sp}.

\begin{figure}[hbt]
    \centering
        \includegraphics[width=0.45\textwidth]{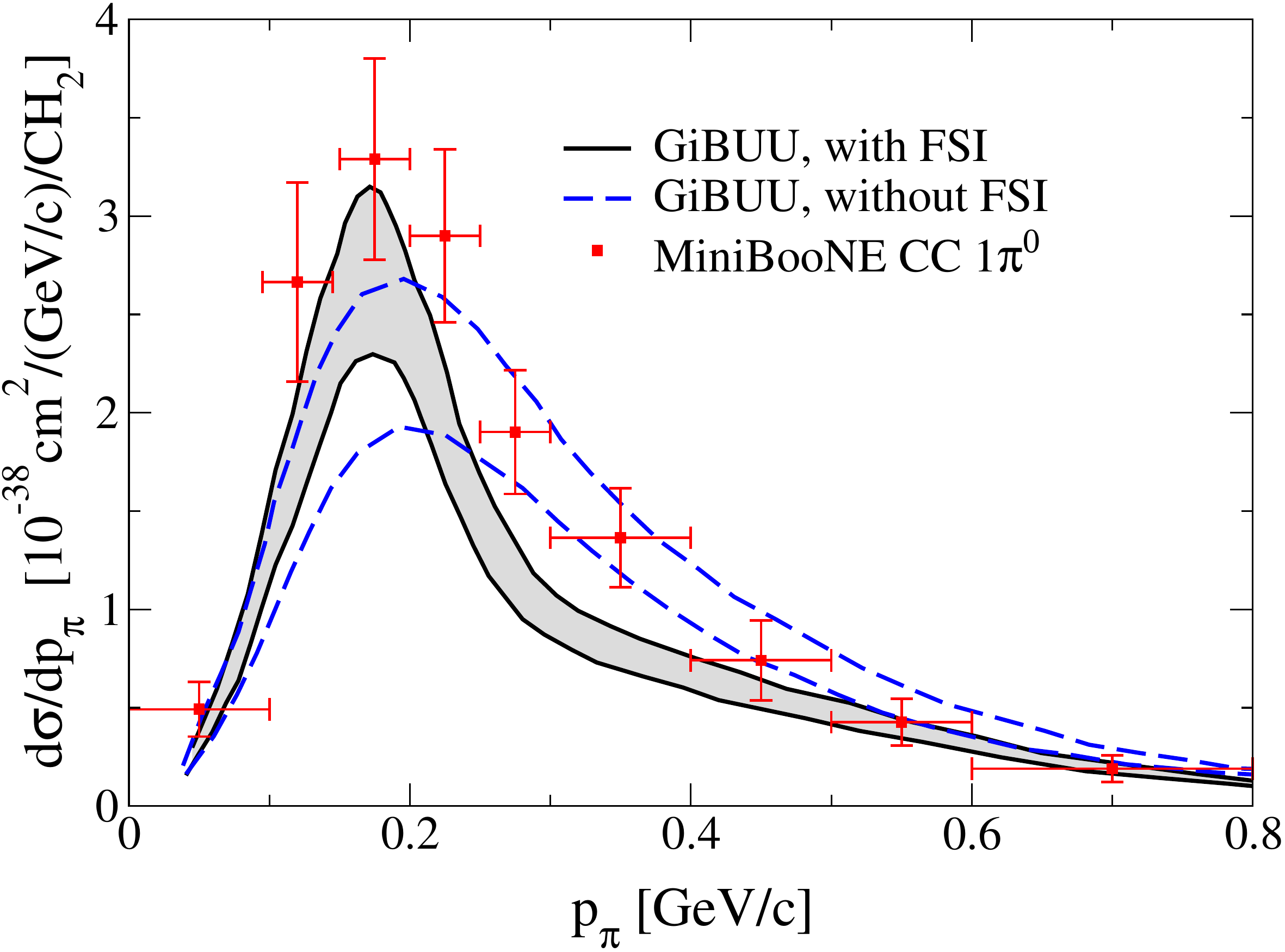}
    ~$ \hspace{0.3cm} $~
        \includegraphics[width=0.45\textwidth]{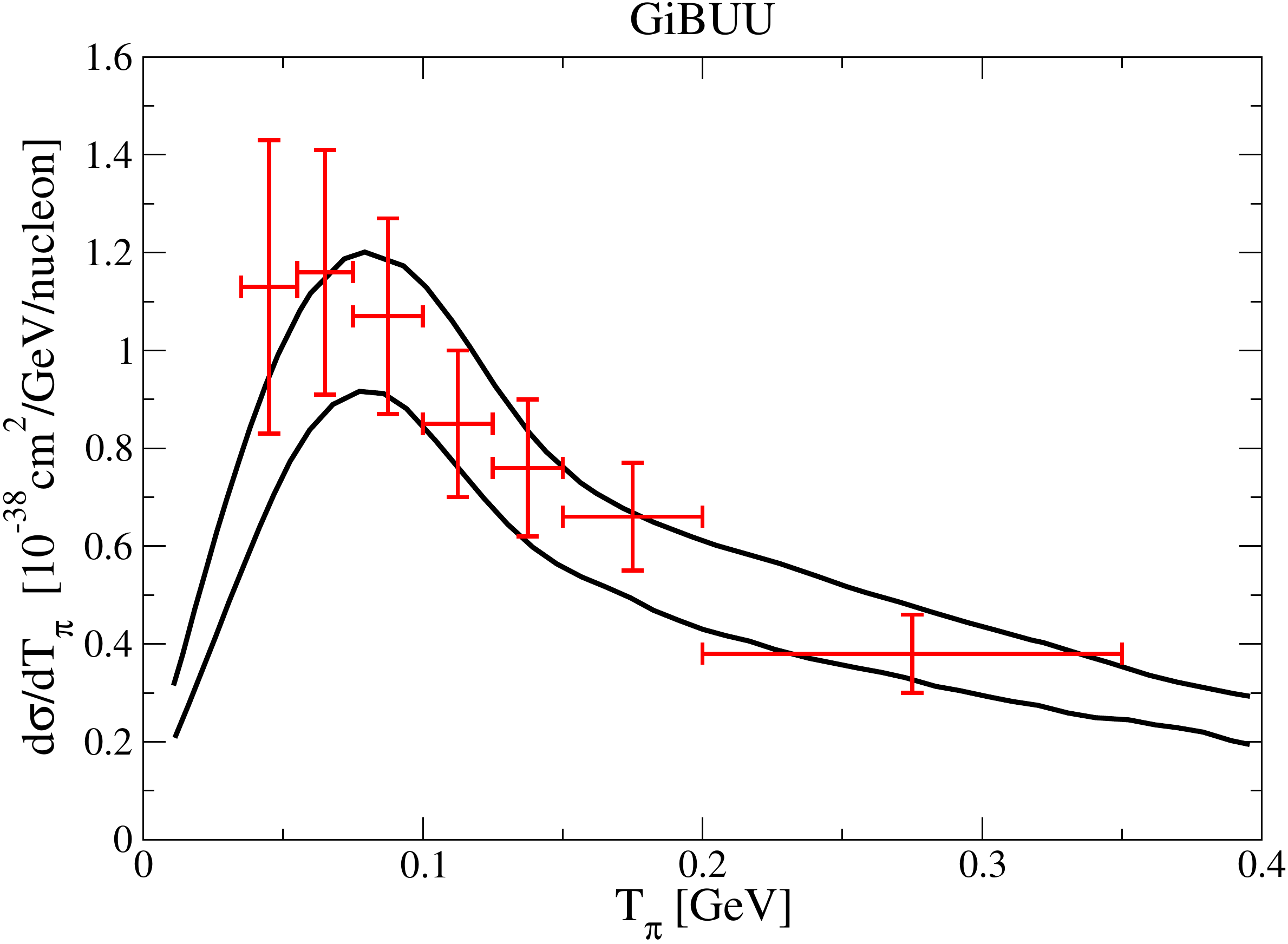}
    \caption{Predictions from the GiBUU transport model for weak pion production. Left: the CC1$\pi^0$ differential cross section on CH$_2$ folded with the $\nu_\mu$ flux at the MiniBooNE detector as a function of the pion momentum~\cite{Lalakulich:2012cj} compared to data from Ref.~\cite{MiniBooNE:2010cxl}. Right:  differential cross section for CC$\pi^\pm$ on CH averaged over the MINERvA low-energy flux as a function of the pion kinetic energy~\cite{Mosel:2015tja} compared to data from Ref.~\cite{MINERvA:2014ogb}.}
    \label{fig:pionspectra}
\end{figure}

In order to minimize the impact of uncertainties and mismodeling on the NC$\pi^0$ background determination, the MiniBooNE experiment relies on its own measurement of the  NC$\pi^0$  reaction~\cite{MiniBooNE:2009dxl} to tune the simulation. One should nonetheless bear in mind that the theoretical description of MiniBooNE pion production data has encountered difficulties.
The left panel of Fig.~\ref{fig:pionspectra} shows that the Giessen Boltzmann-Uehling-Uhlenbeck transport model (GiBUU) fails to reproduce the experimental $\pi^0$ spectrum for pion momenta between 200 and 500 MeV/c~\cite{Lalakulich:2012cj}. The shape disagreement apparent in Fig~\ref{fig:pionspectra} (left) is in contrast with the result of the GiBUU model for the CC$\pi^{\pm}$ (mostly $\pi^+$) reaction compared to MINERvA data. The right panel of Fig~\ref{fig:pionspectra} is adapted from Fig.~1 of Ref.~\cite{Mosel:2015tja}. The band between the two solid lines represent the uncertainty from ANL and BNL data~\cite{Mosel:2015tja}~\footnote{This band would be narrower and closer to the lower end  if the reanalyzed ANL and BNL data of Ref.~\cite{Wilkinson:2014yfa} had been used.}.  

\begin{figure}[h!]
    \centering
    \includegraphics[width=0.65\textwidth]{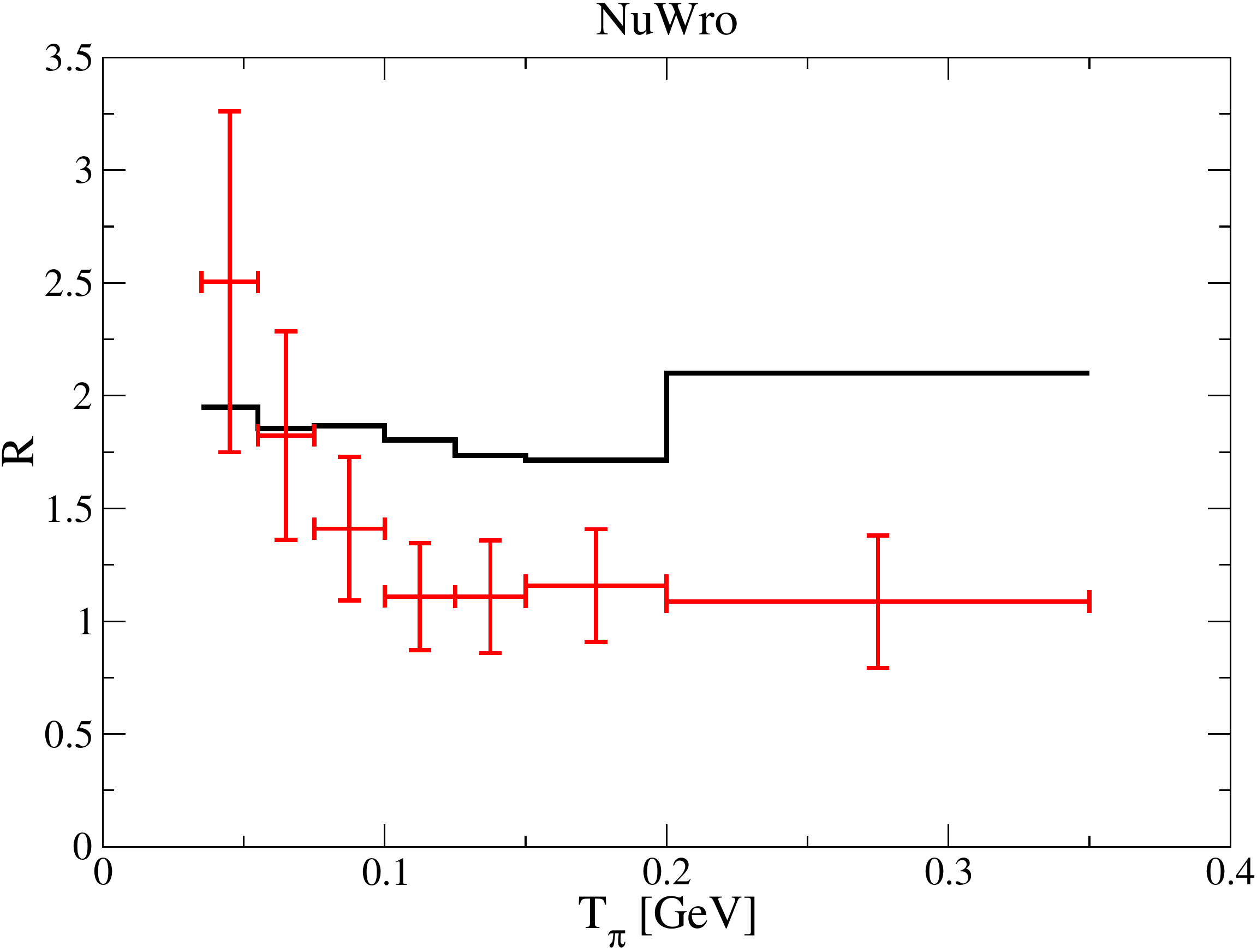}
    \caption{Ratio of $d\sigma/dT_\pi$ from MiniBooNE and MINERvA [Eq.~(\ref{eq:ratio})], and the corresponding NuWro predictions~\cite{Sobczyk:2014xza}.}
    \label{fig:NuWro}
\end{figure}

One is tempted to attribute the different scenarios displayed by Fig~\ref{fig:pionspectra} to the differences in the corresponding neutrino fluxes. The flux at MiniBooNE peaks at nearly 700~MeV (Fig.~\ref{ch:H_nu:sec:MiniBooNE:fig:MiniBooNE_flux}) while the MINERvA low-energy one does close to 3~GeV. However, the GiBUU model also describes well pion production at T2K (Figs.~2 and 3 of Ref.~\cite{Mosel:2017nzk}), whose flux peaks at a around the same energy as the MiniBooNE one. Moreover, according to the study of Ref.~\cite{Sobczyk:2014xza}, there is a strong correlation among the two data sets (at least for charged pions) in spite of the flux differences. Using the NuWro generator, the authors of Ref.~\cite{Sobczyk:2014xza} have obtained that the ratio
\be
R(T_\pi) =\frac{ \left(d\sigma / dT_\pi\right)_{\mathrm{MINERvA,\,CC}\pi^\pm}}
{\left(d\sigma / dT_\pi\right)_{\mathrm{MiniBooNE,\,CC}\pi^+}}  \approx 2
\label{eq:ratio}
\ee
as can be seen in Fig.~\ref{fig:NuWro}, adapted from Ref.~\cite{Sobczyk:2014xza}. In both experiments, the dominant contribution comes from the $\Delta(1232)$ region. The cut in the reconstructed invariant mass $W_\mathrm{rec} \equiv \sqrt{m_N^2 + 2 m_N q_0 + q^2} < 1.4$~GeV is applied in the MINERvA analysis~\cite{MINERvA:2014ogb} using measured lepton kinematics and calorimetry. It quenches the contribution from higher invariant masses although the cut is not sharp, and the $\Delta$ peak is shrunk from its maximum on~\cite{Mosel:2015tja}. As shown in Fig.~\ref{fig:NuWro}, the correlation obtained in Ref.~\cite{Sobczyk:2014xza} with NuWro is absent in the data. 

Even if established specifically for charged pions, this unresolved tension could have implications for the determination of this important background in the MiniBooNE oscillation measurement. Indeed, it would be interesting to study the NC$\pi^0$ background prediction based on a pion production model, like GiBUU, that explains MINERvA data but such an exercise requires a simulation of the MiniBooNE detector. It should be however added that the NC$\pi^0$ background has more events near the edge of the fiducial volume because of the greater chance that one of the decay photons leaves the detector, while electron events are more homogeneously distributed over the detector volume. In Ref.~\cite{MiniBooNE:2020pnu}, the MiniBooNE experiment used this feature of the radial distribution of event vertices to show that an explanation of the anomaly in terms of unaccounted NC$\pi^0$ electron-like events or, in general, due to entering or exiting photons, is disfavored.   

\paragraph{Single-gamma background.} As previously stated, Cherenkov detectors like MiniBooNE misidentify single photon tracks as electrons. Apart from the sources discussed above, such single photons can be produced in NC interactions, NC$1\gamma$, inside the fiducial volume. Although it is not exactly the case (see below), MiniBooNE assumes that the NC$1\gamma$ events come entirely from $\Delta(1232)$ radiative decay: $\Delta \raw N \, \gamma$ and constrains it using the NC$\pi^0$ data~\cite{MiniBooNE:2009dxl} and the NC$1\gamma$/NC$\pi^0$ ratio, taken to be $0.0091 \pm 0.0013$~\cite{MiniBooNE:2020pnu}. The derivation of this number is sketched in Ref.~\cite{MiniBooNE:2020pnu} and reproduced here: the $\Delta(1232)$ contribution to the NC$\pi^0$ event sample is 52.2\% on $^{12}$C and 15.1\% on H$_2$; the $\Delta \raw N \, \pi^0$ fraction is 2/3 and the probability that a pion escapes from $^{12}$C is estimated to be 62.5\%. Finally, the $\Delta$ radiative branching fraction is 0.60\% (0.68\%) on $^{12}$C (H$_2$). Altogether, for CH$_2$ one has $0.151 / (2/3) \times 0.0068 \time 1.5 + 0.522 / (2/3) /0.625 \times  0.0060 = 0.0091$.  The total uncertainty on this ratio is 14.0\% (15.6\%) in neutrino (antineutrino) mode. Estimated in this way, the NC$1\gamma$ one represents the second largest background in the kinematic region where the excess is found. 

The fact that the single-gamma background is only indirectly determined,  while only upper limits are experimentally available~\cite{NOMAD:2011gyy, T2K:2019odo} has stimulated the theoretical activity to model the NC$1\gamma$ reaction on nucleons and nuclei. The following section summarizes these efforts. 

\section{Theoretical description of photon emission in NC interactions in the Standard Model}
\label{sec:ncgamma}

 Photon emission induced by NC interactions can take place on single nucleons
\be
\label{eq:reac_nucleon}
\nu (\bar{\nu})\, N \rightarrow \nu (\bar{\nu})\, \gamma \, N \,,
\ee
and on nuclear targets
\bea
\label{eq:reac_incoh}
\nu (\bar{\nu})\, A &\rightarrow& \nu (\bar{\nu})\, \gamma \, X  \, \\
\label{eq:reac_coh}
\nu (\bar{\nu})\, A &\rightarrow& \nu (\bar{\nu})\, \gamma \, A \,
\eea
via incoherent [Eq.~(\ref{eq:reac_incoh})] or coherent  [Eq.~(\ref{eq:reac_coh})] scattering. 

\bfig[htb!]
\bcen
\includegraphics[width=0.23\textwidth]{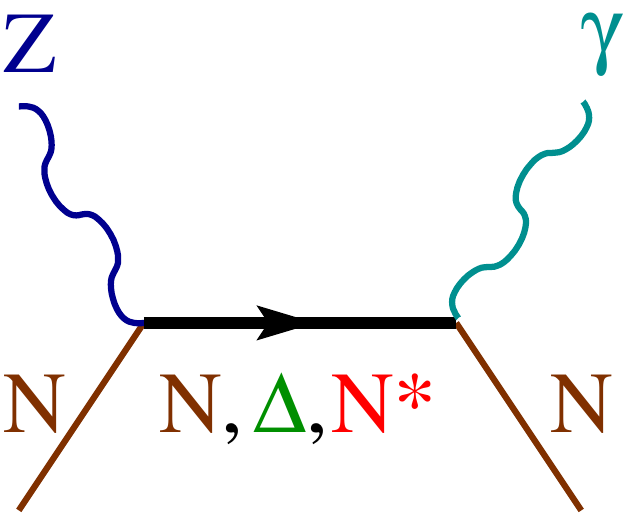}
\hspace{.05\textwidth} 
\includegraphics[width=0.23\textwidth]{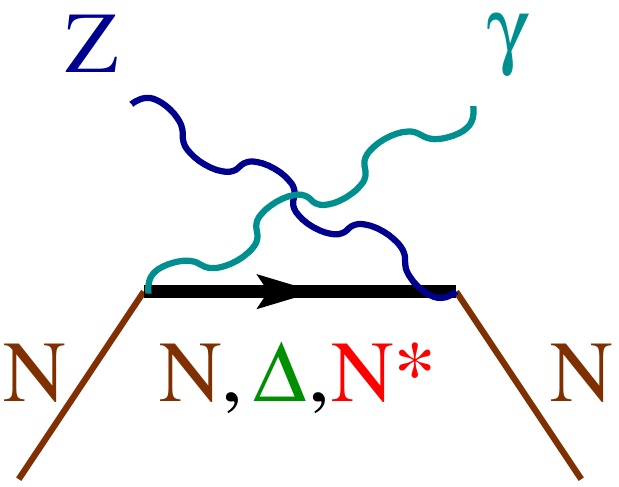}
\hspace{.05\textwidth} 
\includegraphics[width=0.2\textwidth]{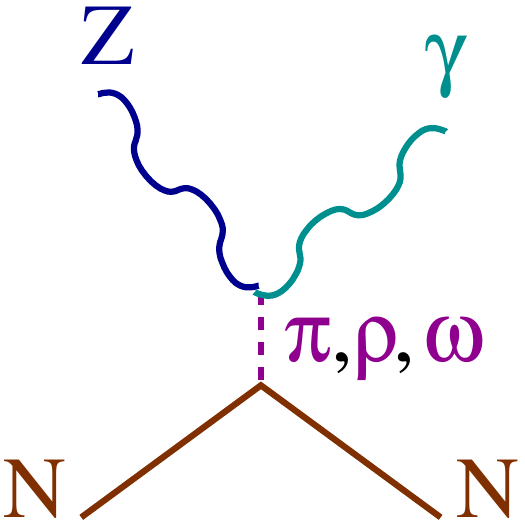}
\ecen
\caption{\label{fig:NCgamma_diags}  Feynman diagrams for NC photon emission considered in the literature. The first two diagrams stand for direct and crossed baryon pole terms with nucleons and baryon resonances $\Delta(1232)$, $N^*(1440)$, $N^*(1520)$, $N^*(1535)$ in the intermediate state. The third diagram represents $t$-channel meson ($\pi$, $\rho$, $\omega$) exchange contributions.}
\efig
Theoretical models for the reaction of Eq.~(\ref{eq:reac_nucleon}) in the few-GeV region have been developed in Refs.~\cite{Hill:2009ek,Serot:2012rd,Wang:2013wva}. These calculations incorporate $s$- and $u$-channel amplitudes with nucleons and $\Delta(1232)$ in the intermediate state, Fig.~\ref{fig:NCgamma_diags}. The structure of nucleon pole terms at threshold is fully determined by the symmetries of the SM. The extension towards higher energy and momentum transfers, required to predict cross sections at MiniBooNE, is performed by the introduction of phenomenologically parametrized  weak and electromagnetic form factors. The same strategy has been followed for the resonance terms. 
\begin{figure}[h!]
    \centering
        \includegraphics[width=0.48\textwidth]{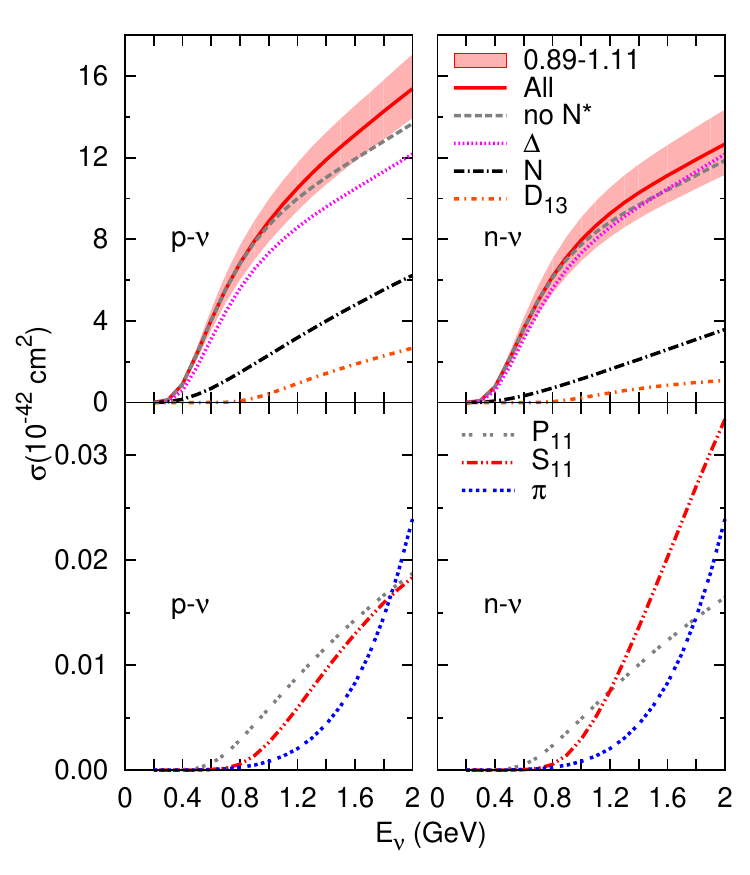}
        \includegraphics[width=0.48\textwidth]{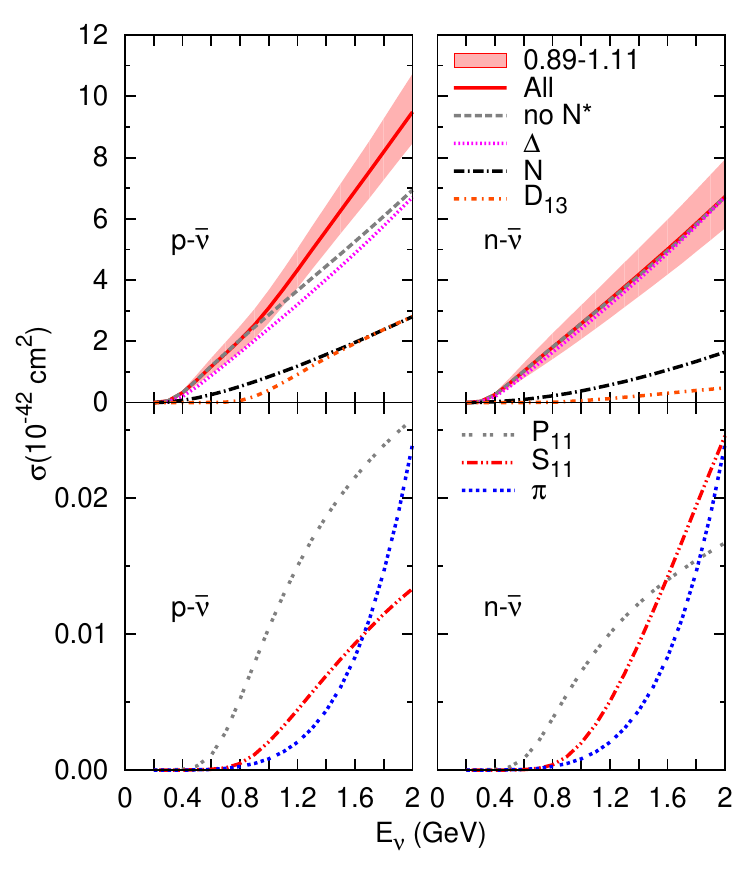}
    \caption{NC$1\gamma$ cross sections on protons and neutrons as a function of the (anti)neutrino energy according to the model of Ref.~\cite{Wang:2013wva}. The error bands in the full-model results (solid lines) represent the uncertainty in the leading axial $N \Delta$ coupling denoted as $C_5^A(0)$.}
    \label{fig:ncgammacs}
\end{figure}
The cross sections on elementary targets obtained in Ref.~\cite{Wang:2013wva} are reproduced in Fig.~\ref{fig:ncgammacs}. They show that the $\Delta(1232)$ excitation followed by radiative decay is the dominant mechanism, as correctly assumed by MiniBooNE, and in agreement  with the findings of Refs.~\cite{Hill:2009ek,Serot:2012rd}. Nonetheless, the contribution from the $N(1520)3/2^-$ on proton targets is sizable above $E_\nu \sim 1.5$~GeV, while $N(1440)1/2^+$ and $N(1535)1/2^-$ are negligible. The pion-pole mechanism, which originates from the $Z^0\gamma\pi$ vertex, fixed by the axial anomaly of QCD, is nominally of higher order~\cite{Serot:2012rd} and, indeed, gives a very small contribution to the cross section. Among terms with heavier meson $t$-channel exchange, the $\omega$ one was proposed as a solution for the MiniBooNE anomaly~\cite{Harvey:2007rd} because of the rather large (although uncertain) couplings and the $\omega$ isoscalar nature, which enhances its impact on the coherent reaction of Eq.~(\ref{eq:reac_coh}). However, actual calculations found this contribution small compared to $\Delta(1232)$ excitation~\cite{Hill:2009ek,Rosner:2015fwa}.

\begin{figure}[h!]
    \centering
    \includegraphics[width=0.5\textwidth]{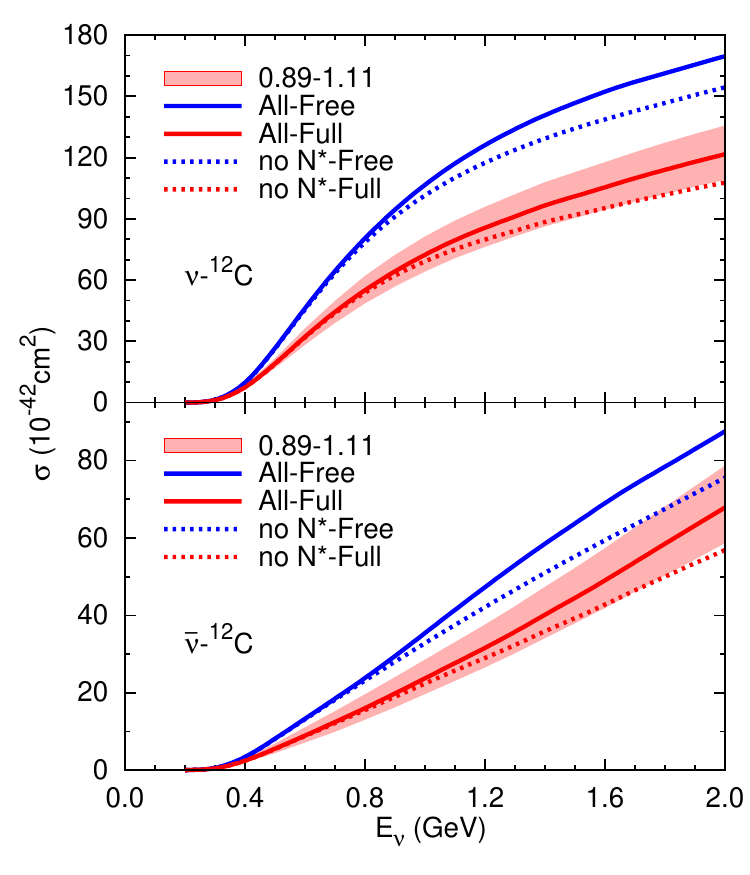}
    \caption{Neutrino (top) and antineutrino (bottom)
   incoherent photon emission cross sections on $^{12}$C.  Curves denoted as ``Free'' do not include any nuclear correction while those labeled as ``Full'' take into account Pauli blocking, Fermi motion and the in medium $\Delta$ resonance
  broadening. The error bands show the uncertainty on the full model from $C_5^A(0)$.}
    \label{fig:ncgammanuclearcs}
\end{figure}
The incoherent NC$\gamma$ reaction on nuclear targets, Eq.~(\ref{eq:reac_incoh}), has been studied in Refs.~\cite{Zhang:2012aka,Wang:2013wva} using the relativistic local Fermi gas approximation to take into account Fermi motion and Pauli blocking. The broadening of the $\Delta$ resonance in the medium has also been  incorporated using a spreading potential in Refs.~\cite{Zhang:2012aka,Zhang:2012xn}, while Ref.~\cite{Wang:2013wva} uses the parametrization of the imaginary part of the in-medium $\Delta$ selfenergy as a function of the local nuclear density derived in Ref.~\cite{Oset:1987re}. The  neglect of nuclear medium corrections is a poor approximation: by taking into account Fermi motion and Pauli blocking, the cross section already goes down by more than 10\%. With the full model the reduction is  of the order of 30\% as can be seen in Fig.~\ref{fig:ncgammanuclearcs} taken from Ref.~\cite{Wang:2013wva}).

\begin{figure}[h!]
    \centering
        \includegraphics[width=0.48\textwidth]{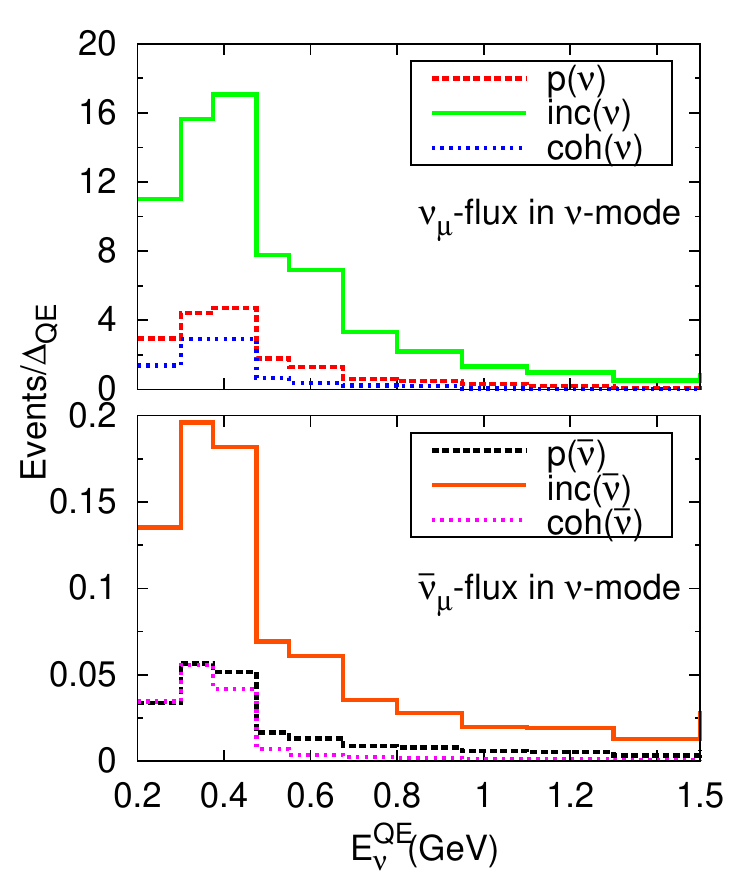}
        \includegraphics[width=0.48\textwidth]{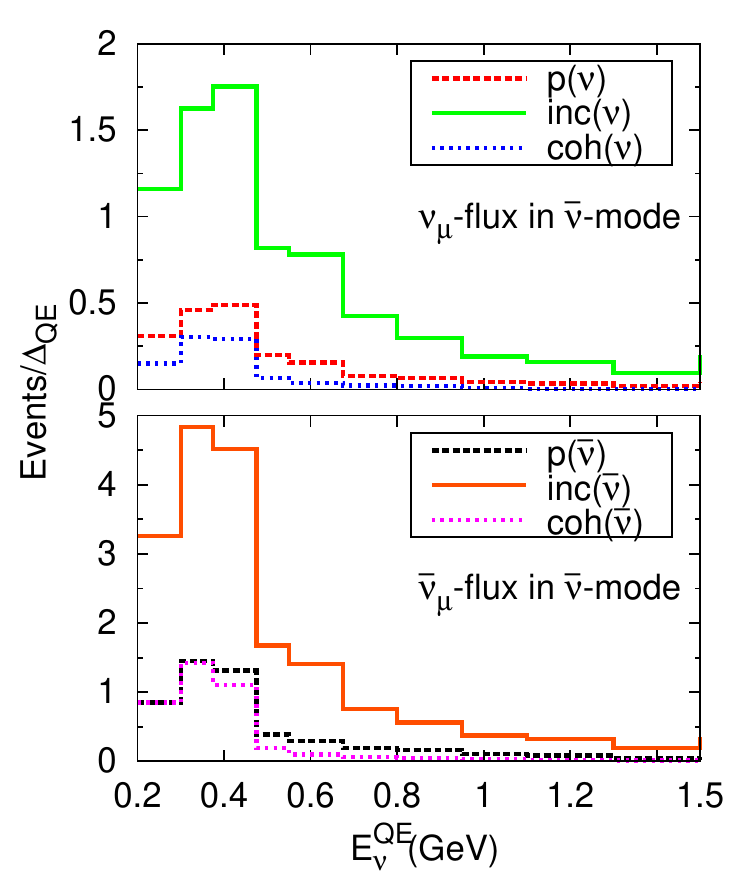}
    \caption{Predicted distribution of NC$1\gamma$ events at MiniBooNE as a function of $E^\mathrm{QE}_{\nu}$ for the $\nu_\mu$ (top)  and $\bar \nu_\mu$  (bottom) MiniBooNE fluxes in the $\nu$(left) and $\bar \nu$ (right) modes~\cite{Wang:2014nat} calculated with the model of Ref.~\cite{Wang:2013wva}.}
    \label{fig:photonevents}
\end{figure}

The NC$\gamma$ model outlined above has been applied to calculate the number and distributions of single photon events at MiniBooNE~\cite{Wang:2014nat}, using the available information about the detector mass and its composition (CH$_2$), the  number of POT,~\cite{Aguilar-Arevalo:2013pmq}, flux prediction (Fig.~\ref{ch:H_nu:sec:MiniBooNE:fig:MiniBooNE_flux}) and photon detection efficiency~\cite{webpage}. As shown in Fig.~\ref{fig:photonevents} taken from Ref.~\cite{Wang:2014nat}, yields from the incoherent channel are the largest ones. Those from the coherent channel and the reaction on protons, which are comparable, are smaller but significant. 

\begin{figure}[h!]
\bcen  
\includegraphics[width=0.45\textwidth]{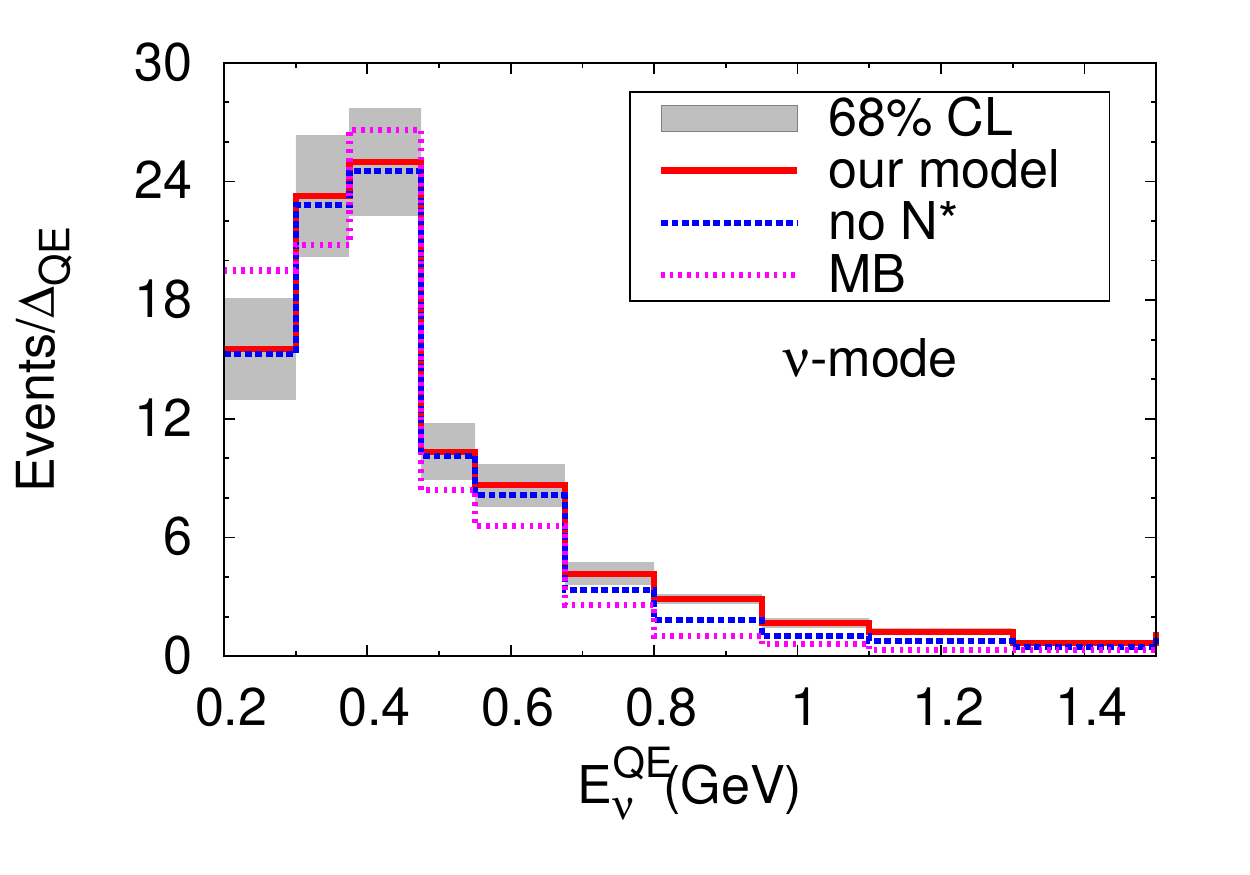}
\includegraphics[width=0.45\textwidth]{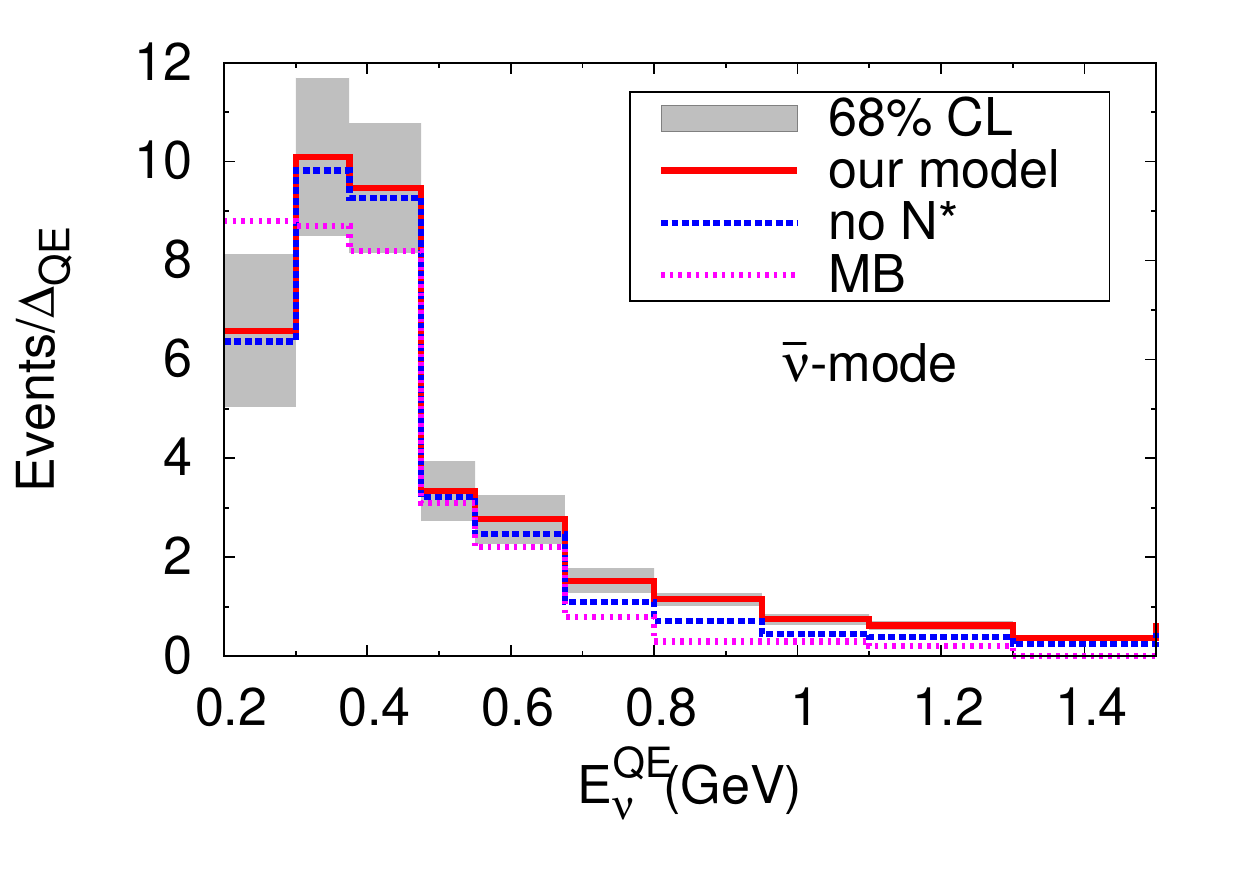}  
\caption{
\label{fig:re}
 $E^\mathrm{QE}_\nu$ distributions of total  NC$\gamma$  events 
for the $\nu$ (left) and $\bar \nu$ (right) modes. The ``MB'' histograms display the MiniBooNE estimates~\cite{webpage}.}
\ecen
\end{figure}

The sum of all contributions in Fig.~\ref{fig:photonevents} are shown in Fig. ~\ref{fig:re}. The error bands correspond to a 68~\% confidence level according to the error budget in Table~1 of Ref.~\cite{Wang:2014nat} and is dominated by the uncertainty in the $C_5^A(0)$ $N\Delta$ axial coupling. The comparison with the MiniBooNE estimate described above shows a good agreement: the shapes are similar and the peak positions coincide. The largest discrepancy is observed in the lowest energy bin. In the two bins with the largest number of events, the two calculations are consistent within our errorbars. The overall agreement is also good in comparison to the result of Zhang and Serot~\cite{Zhang:2012xn}.

 We should warn the reader that all the results shown in this section were obtained assuming a value of $C^A_5(0) = 1.00 \pm 0.11$ determined in a fit to $\nu_\mu d \to \mu^- p \, \pi^+ \, n$ BNL and ANL data in the $\Delta$ region~\cite{Hernandez:2010bx}. However, a re-analysis of these data with an improved version of the weak pion production model obtained a higher $C^A_5(0) = 1.18 \pm 0.07$, which is also in excellent agreement with the off-diagonal Goldberger-Treiman relation for the $N\Delta$ transition. This change would cause an increase in the NC$1\gamma$ cross section and in the number of events predicted at MiniBooNE, leading to a better agreement with the MiniBooNE determination in the first bin, but overestimating it in most bins. This is, after all, not surprising given the fact that theory takes into account NC$1\gamma$ mechanisms that are unaccounted by MiniBooNE. In any case, such an increase would be  insufficient to explain the anomaly. The same conclusion applies to the study of Ref.~\cite{Chanfray:2021wfw} where a two-nucleon $\Delta$ mediated meson exchange mechanism was considered for the average of NC neutrino and antineutrino cross sections. Its contribution appears to be significant but, as expected, smaller (a factor of around 9 at $E_\nu=500$~MeV) than the single nucleon mechanisms discussed above. 

As stated above, the coherent reaction channel, Eq.~\ref{eq:reac_coh}, is responsible for a small but non-negligible fraction of the NC$1\gamma$ events at MiniBooNE. It is particularly important for antineutrinos and in the forward direction  (see Fig.~3 of Ref.~\cite{Wang:2014nat}). Furthermore, coherent NC$1\gamma$ emission appears as a background to some of the proposed explanations of the MiniBooNE event excess outlined in Sec.~\ref{sec:expl},  which involve physics beyond the SM and can be tested at Fermilab by the Short Baseline Neutrino Detector (SBND) or by MINERvA. For these reasons we dwell longer on the theoretical description of this process. 

\paragraph{Coherent photon emission in NC interactions}

\begin{figure}[h!]
    \centering
    \includegraphics[]{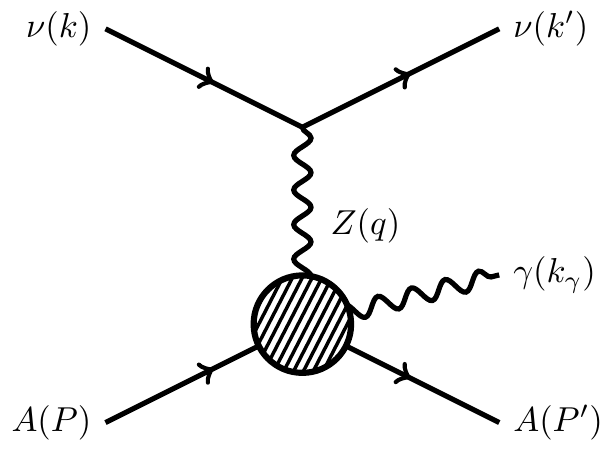}
    \caption{Diagram for coherent photon emission induced by neutral currents.}
    \label{fig:cohdiag}
\end{figure}

In the process of Eq.~\ref{eq:reac_coh}, diagrammatically illustrated in Fig.~\ref{fig:cohdiag}, a neutrino with four-momentum $k \equiv ( E_\nu, \vec{k} )$ interacts with a nucleus of four-momentum $P \equiv \left( E, \vec{P} \right)$.
After the interaction, the nucleus remains in the ground state, changing only its four-momentum to  $P' \equiv \left( E', {\vec{P}\,}' \right)$, while the neutrino does to $k' \equiv ( E'_\nu, {\vec{k}\,}' )$. 
The four-momentum of the emitted photon is $k_\gamma \equiv \left( E_\gamma, \vec{k}_\gamma \right)$ and the one transferred by the neutrino is $q = k - k'$. In the Laboratory frame $P = \left(M_A, 0 \right)$, where $M_A$ denotes the target mass. Under the assumption that the recoil kinetic energy of the outgoing nucleus $(E'-M_A) \ll M_A$, energy conservation implies that $q_0 = E_\gamma$.

In the framework adopted here, which is adapted from Refs.~\cite{Alvarez-Ruso:2007rcg,Amaro:2008hd} for neutrino-induced coherent pion production reactions, the cross section in the Laboratory frame is given by 
\begin{equation}
    \frac{d\sigma}{dE_\gamma d\Omega_\gamma d\Omega_{k'}} = \frac{1}{8} \frac{1}{\left(2\pi\right)^5} \frac{E'_\nu E_\gamma}{E_\nu} |\overline{\mathcal{M}}|^2
    \:.
    \label{ch:photoproduction:sec:cross_section:eq:diff_CS}
\end{equation}
Owing to the coherence, the nucleon wave function inside the nucleus remains unchanged. Hence, after summing the elementary $Z N \rightarrow N \gamma$  amplitudes over all nucleons, one obtains the nuclear density distributions of protons and neutrons, $\rho_{p,n}(r)$. The absolute value  of the amplitude squared can be cast as 
\begin{equation}
    |\overline{\mathcal{M}}|^2 = - \frac{G_F^2}{2} e^2 L_{\alpha \beta} \, g_{\mu \nu} \left( R^{\mu \alpha} \right)^\dagger R^{\nu \beta}
    \:,
\end{equation}
where 
\begin{equation}
    L_{\mu\nu} = 8\left[k'_\mu \, k_\nu + k'_\nu \, k_\mu - g_{\mu\nu} (k'\cdot k) \pm i \epsilon^{\mu\nu\sigma\rho} k'_\sigma k_\rho \right]
\end{equation}
is the standard leptonic tensor, and
\begin{equation}
    \begin{aligned}
        R^{\mu\alpha} = \sum_{N=p,n} \frac{1}{2} \text{Tr} \left[ \frac{\slashed{p} + m_N}{2 m_N} \gamma_0 \Gamma_N^{\mu \alpha} \right] \frac{m_N}{E_p} F_N \left(\left| \vec{q}-\vec{k}_\gamma  \right| \right)
        \:,
    \end{aligned}
    \label{eq:Hadronic_current_R_with_FF}
\end{equation}
in terms of the amputated amplitudes $\Gamma_N^{\mu \alpha}$  corresponding to the $Z\, N \raw N \, \gamma$ matrix elements: 
\begin{equation}
    \bra{N \gamma} J^\mu_{\text{NC}\gamma} \ket{Z N} = \overline{u} \left( p' \right) \Gamma_N^{\mu \alpha} u \left( p \right) \epsilon_\alpha^* \left( k_\gamma \right)
    \:,
\end{equation}
with $\epsilon_\alpha (k_\gamma)$ the polarization of the outgoing photon. The nuclear form factors $F_N$ arise as the Fourier transform of $\rho_{p,n}(r)$ 
\begin{equation}
F_N \left(\left| \vec{q}-\vec{k}_\gamma  \right| \right) = \int d^3r \, e^{i \left(\vec{q}-\vec{k}_\gamma \right)\vec{r}} \, \rho_N(\vec{r})   
\end{equation}
once the amputated amplitudes are evaluated at the average nuclear density and factorized out of the integration over $\vec{r}$. For the so far undefined nucleon momenta, we assume that the momentum transferred to the nucleus is equally shared by  initial and final on-shell nucleons, so that  $p = \left( E_p, \vec{p}\right)$ with $\vec{p} = (\vec{k}_\gamma - \vec{q})/2$ and $E_p = \sqrt{m_N^2 + {\vec{p}\,}^2}$. This prescription is justified in Ref.~\cite{Amaro:2008hd}, where it is also shown how the sum over nucleon helicities together with the choice of nucleon momenta leads to the trace in Eq.~\ref{eq:Hadronic_current_R_with_FF}.

The elementary $Z N \rightarrow N \gamma$ amplitude includes the same mechanisms discussed above and represented by the Feynman diagrams of Fig.~\ref{fig:NCgamma_diags}.
In the case of the coherent process, nucleon-pole contributions are negligible;
$\pi$ and $\rho$ exchange terms are not only small but, in the coherent case, vanish exactly for isospin symmetric nuclei.
The $\omega$ exchange contribution, instead, does not vanish for symmetric nuclei because amplitudes on protons and neutrons add up rather than subtract. This mechanism was found subdominant at $E_\nu \sim 1$~GeV~\cite{Hill:2009ek,Zhang:2012xi}. Its relevance at higher energies is highly uncertain due to a high sensitivity to unknown form factors and unitarity constraints but cannot be discarded, due to its strong energy dependence~\cite{Zhang:2012xi}, and deserves future studies. Here we focus on the contribution from baryon-resonance ($N^*$ and $\Delta$) intermediate states. The calculation of Ref.~\cite{Wang:2013wva} considered $\Delta(1232)$, $N(1440)$, $N(1520)$ and $N(1535)$ intermediate states. Keeping in mind that there are experiments which work with higher energy fluxes, like MINER$\nu$A, where the medium-energy flux peaks at around 6 GeV and can detect photons with energies above 500~MeV, we have extended the validity of the model to this domain~\cite{Edutesis}. This is done by adding to the amplitude new resonant diagrams for all the $N^*$ and $\Delta$ states with invariant masses $W < 2$~GeV whose electromagnetic helicity amplitudes, upon which we rely, were extracted with the Mainz Unitary Isobar model (MAID)~\cite{Drechsel:2007if,Tiator:2011pw}. 

Nuclear medium modifications of the resonant elementary amplitudes are neglected for all the states except the $\Delta(1232)$, which dominates the cross section and is known to be strongly modified in this medium. This is done by changing the $\Delta \raw N \, \pi$ decay width to account for the Pauli blocking of the final nucleon and introducing an average broadening of twice the spreading potential $V_0 = 80$~MeV~\cite{Zhang:2012aka}. The obtained results are consistent with those obtained with the more sophisticated self-energy of Ref.~\cite{Oset:1987re}.

\begin{figure}[h!]
\bcen  
\includegraphics[width=0.49\textwidth]{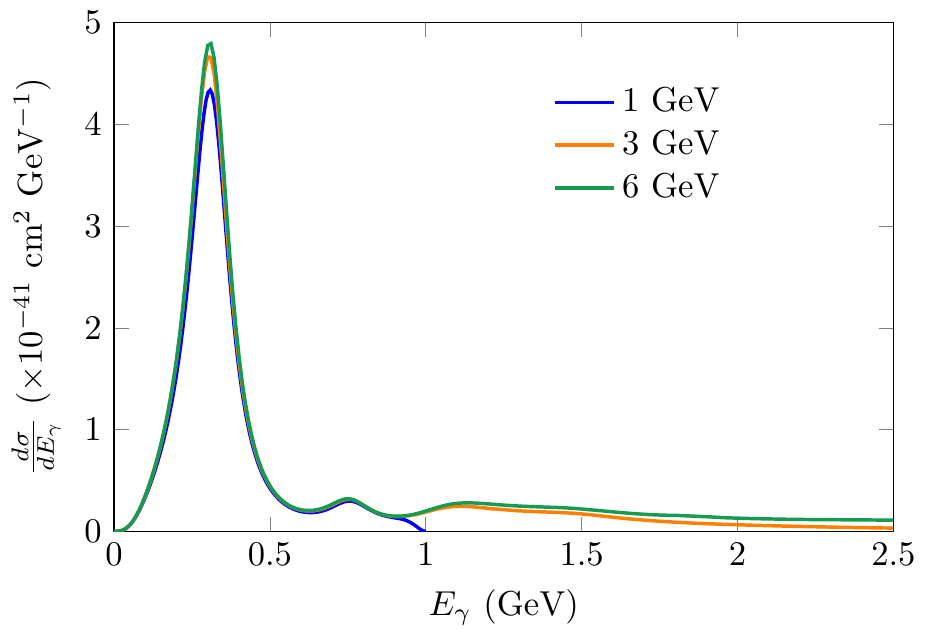}
\includegraphics[width=0.47\textwidth]{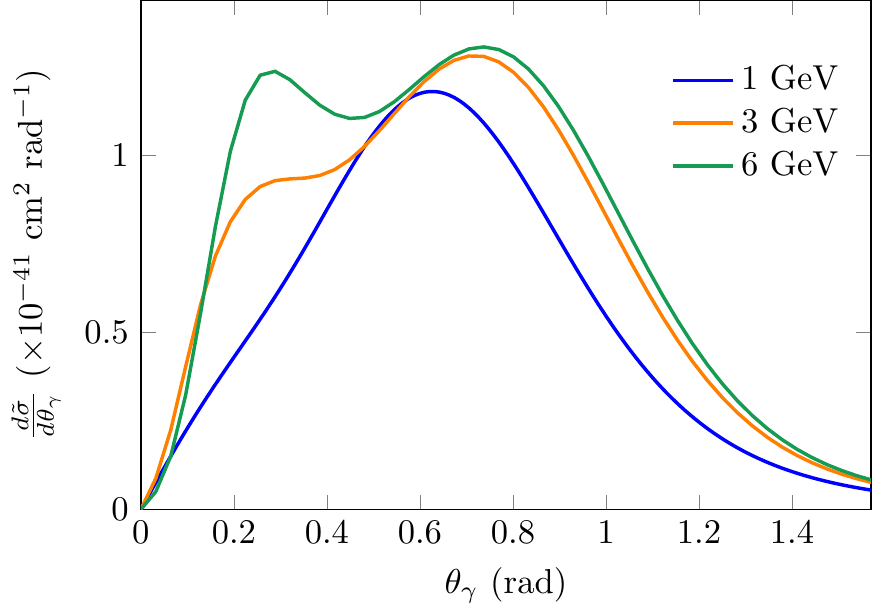}  
\caption{Photon energy (left) and angular (right) distributions for coherent NC$\gamma$ on $^{12}$C for 1 GeV, 3 GeV and 6 GeV incoming neutrinos. For the angular distributions, only photon energies $E_\gamma < 2.5$~GeV have been considered.}
 \label{fig:Sum_all_6GeV}
\ecen
\end{figure}

In Fig~\ref{fig:Sum_all_6GeV} (left) the $E_\gamma$ distributions are shown at 1, 3 and 6 GeV of incoming neutrino energy. The tendency of the cross section towards saturation is apparent, with small differences between the results at 3 and 6 GeV. These plots clearly show the dominant role of the $\Delta (1232)$. Some strength comes also from  $N(1520)$ for $E_\gamma < 1$~GeV. For $E_\gamma > 1$~GeV, several resonances overlap but the only non negligible strength is provided by $\Delta(1700)$ and $\Delta(1950)$.
Angular distributions, Fig.~\ref{fig:Sum_all_6GeV} (right),  are forward peaked as expected for coherent scattering. As the neutrino energy increases, more strength is accumulated at small photon angles. For high energy photons, a small change in the angle will be highly disfavored by the nuclear form factor. As a  consequence, the forward peak at around 0.2~rad in Fig.~\ref{fig:Sum_all_6GeV} (right) is a reflection of the contribution of the high energy tail in the energy distribution.

\section{Some possible explanations of the anomaly.}
\label{sec:expl}

\paragraph{Neutrino oscillations.} As anticipated in the introduction, the main hypothesis to explain short baseline anomalies has traditionally been, and remains to be, the existence of additional families of sterile neutrinos able to mix with the SM ones.  
In a combined fit within the two-neutrino oscillation model to the final results in both neutrino and antineutrino mode, MiniBooNE finds a good description of the data. The best-fit point $(\sin^2{2 \theta}, \Delta m^2 ) = (0.807, 0.043 \, \mathrm{eV}^2)$ is favored with respect to the background-only fit, which has a $3 \times 10^{-7}$ smaller $\chi^2$ probability. 

\begin{figure}[h!]
\bcen  
\includegraphics[width=0.48\textwidth]{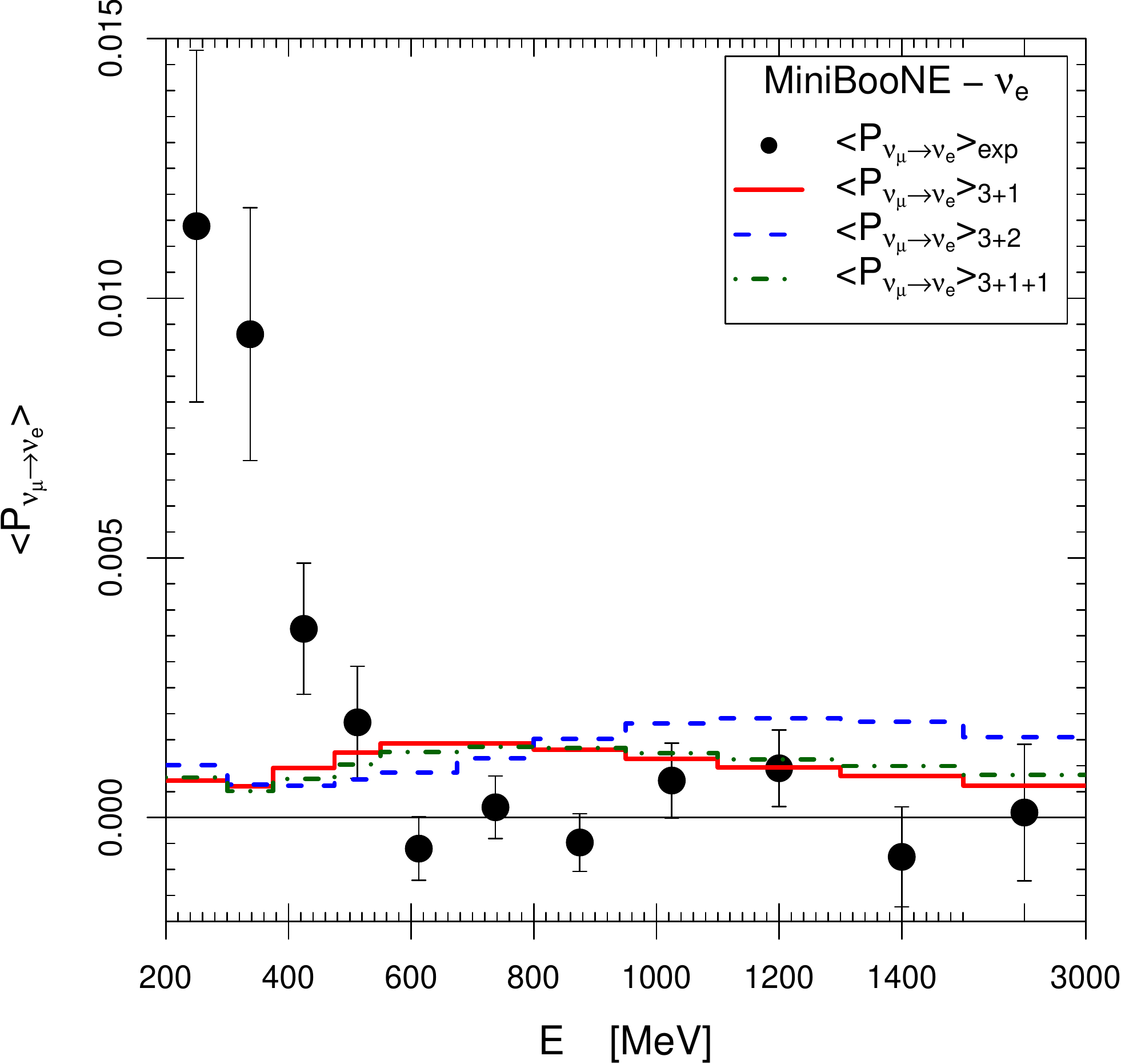}
\includegraphics[width=0.48\textwidth]{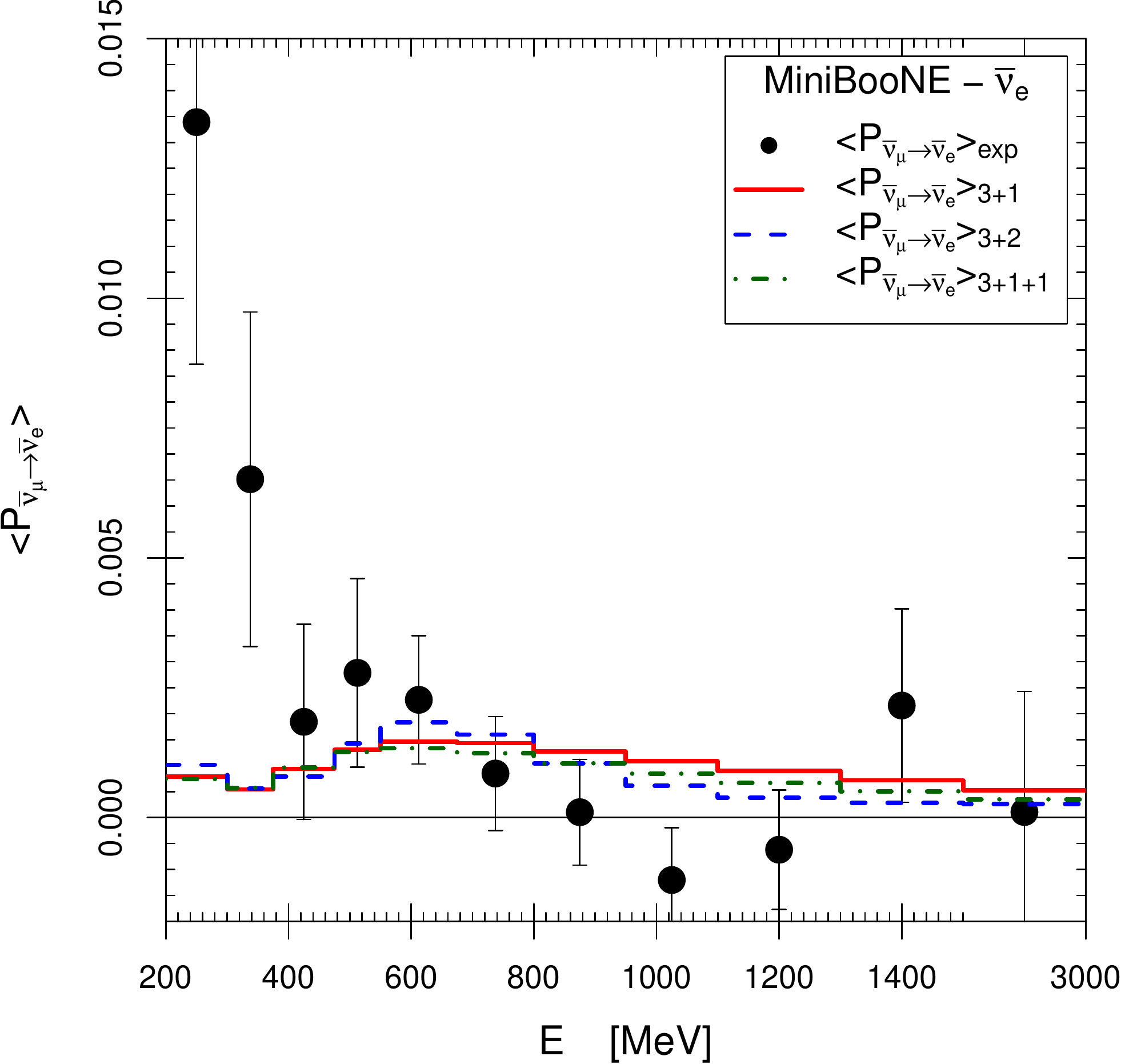}  
\caption{Averaged transition probability in neutrino-energy bins corresponding to the best-fit values of the oscillation parameters in the 3+1, 3+2 and 3+1+1 fits~\cite{Giunti:2013aea} compared to the MiniBooNE excess data.}
 \label{fig:Giunti}
\ecen
\end{figure}

On the other hand, the MiniBooNE results should be considered in the global context of other limits and signals. It was early observed that  a $3+1$ model with three active and one sterile neutrinos offers a poor compatibility between $\nu$ and $\bar\nu$ datasets and an even poorer one (0.0013\%) between appearance and disappearance measurements~\cite{Conrad:2013mka}. A more recent study~\cite{Dentler:2018sju} also finds a strong tension between appearance (LSND and MiniBooNE) and disappearance (MINOS, IceCube) data, although in this case the tension is dominated by LSND. Models with more sterile-neutrino families offer more flexibility but, nonetheless, global analyses struggle to accommodate the MiniBooNE excess with the world oscillation data even in 3+2 and 3+1+1~\footnote{In the 3+1+1 scheme, the fifth neutrino is much heavier than 1~eV so that oscillations due to $\Delta m^2_{51}$ are averaged~\cite{Giunti:2013aea}.} neutrino mixing schemes, as clearly seen in Fig.~\ref{fig:Giunti} taken from Ref.~\cite{Giunti:2013aea}.  
As discussed in Secs.~\ref{sec:theano} and \ref{sec:thebackg}, multinucleon interactions have an impact on the intrinsic $\nu_e \, (\bar\nu_e)$ background (Fig.~\ref{fig:nue_background}) and introduce a bias in neutrino energy reconstruction. The consequences for global oscillation analyses were investigated in Ref.~\cite{Ericson:2016yjn} with a 3+1 model. It was found that taking multinucleon interactions into account decreases the appearance-disappearance tension but is insufficient to remove it. In Ref.~\cite{Giunti:2019sag} it was estimated that the NC$1\gamma$ background should be enhanced by a factor between 1.52 and 1.62 over the MiniBooNE estimate, depending on the energy range and mode. Such an enhancement, shrinks the excess and significantly reduces the appearance-disappearance tension in global fits but is at odds with the theoretical calculations of the NC$1\gamma$ number of events at MiniBooNE described in Sec.~\ref{sec:ncgamma}. The upper limit for the NC1$\gamma$ cross section recently obtained by the MicroBooNE experiment \cite{MicroBooNE:2021zai } disfavors that the excess could be attributed to this reaction channel alone.  The study of Ref. \cite{ Brdar:2021cgb } finds that the choice of event generator (NUANCE, GiBUU, GENIE, NuWro) has an impact on NC$\pi^0$ and NC1$\gamma$ backgrounds even when they are constrained by MiniBooNE’s own data. The investigation of how the choice of event generator affects the fit in a $3+1$ model concludes that, even in the most favorable scenario, they seem unable to account for the anomaly. 

Exotic mechanisms that can alter oscillations and eventually reconcile appearance and disappearance data have been proposed, and include Lorentz and CPT violation~\cite{Katori:2006mz, Diaz:2011ia}, non-standard interactions~\cite{Liao:2016reh}, sterile neutrinos with modified dispersion relations~\cite{Barenboim:2019hso}, or a combination of oscillations and sterile neutrino decays~\cite{Moulai:2019gpi, Dentler:2019dhz, deGouvea:2019qre, Vergani:2021tgc}. In the following we consider some scenarios directly involving unconventional mechanisms of neutrino interactions at the detector leading to electron-like signals. The framework for global analyses of this kind of models has been presented in Ref.~\cite{Brdar:2020tle}.

\paragraph{Production and radiative decay of heavy neutrinos.}

In an early study, Gninenko proposed that additional photons could originate in the weak production of a heavy ($m_h \approx 50$~MeV) sterile neutrino slightly mixed with muon neutrinos, followed by its radiative decay~\cite{Gninenko:2009ks}. Following that,  it was pointed out in Ref.~\cite{Masip:2012ke} that the $\nu_h$ could also be electromagnetically produced, alleviating tensions in the original proposal with other data such as those from radiative muon capture measured at TRIUMF.

The scenario presented in Ref.~\cite{Masip:2012ke} has been revisited using present understanding of electromagnetic (EM) and weak interactions on nucleons and nuclei~\cite{Edutesis}. The relevant processes are 
\bea
\label{nuN}
\nu_\mu \,, \bar{\nu}_\mu(k) \,+ \,  N(p) &\rightarrow&  \nu_h \,, \bar{\nu}_h(k') \,+\,  N(p') \,,  \\ [.1cm]
\label{nuA}
\nu_\mu \,, \bar{\nu}_\mu(k) \,+\, A(p) &\rightarrow&  \nu_h, \bar{\nu}_h(k')  \,+\, A(p')\,,  \\ [0.1cm]
\label{nuX}
\nu_\mu \,, \bar{\nu}_\mu(k) \,+\, A(p) &\rightarrow&  \nu_h, \bar{\nu}_h(k')  \,+\, X(p')  \,,
\eea
followed by the decay of the heavy neutrino into a photon and a light neutrino, which could or could not be one of the SM flavors. 
Reaction (\ref{nuA}) is coherent while (\ref{nuX}) is incoherent; excited states $X$ include any number of knocked out nucleons  but no meson production. In the MiniBooNE case, the relevant targets are $N=$~proton and $A=^{12}$C (CH$_2$). 

In the case of an EM reaction in which an incoming light neutrino of flavor $i$ turns into an outgoing heavy neutrino by single-photon exchange through a transition magnetic moment  $\mu_\text{tr}^i$, the most general effective interaction~\cite{Broggini:2012df} leads to the effective Lagrangian proposed in Ref.~\cite{Masip:2012ke}
\begin{equation}
    \mathcal{L}_{\text{eff}} \supset \frac{1}{2}\mu_\text{tr}^i\left[\overline{\nu}_h \sigma_{\mu\nu}\left(1-\gamma_5\right)\nu_i+ \overline{\nu}_i \sigma_{\mu\nu}\left(1+\gamma_5\right)\nu_h\right]\partial^\mu A^\nu
    \:.
    \label{eq:Masip_L}
\end{equation}
In this scenario, the heavy neutrinos are Dirac particles with $m_{\nu_h} \gg m_{\nu_i}$. In the weak case, the neutrino vertex has the same structure as in the SM and is proportional to the mixing $U_{\mu h}$. The EM (NC) hadronic tensors are the same probed in the corresponding electron (neutrino) scattering processes. For the nucleon, it is given in terms of electromagnetic and axial form factors. For coherent scattering (\ref{nuA}), the tensor is proportional to the square of the nuclear form factor. The incoherent reaction can be described with particle-hole excitations in infinite nuclear matter, adapted to finite nuclei using the local density approximation. 

\begin{figure}[h!]
    \centering
        \includegraphics[width=0.45\textwidth]{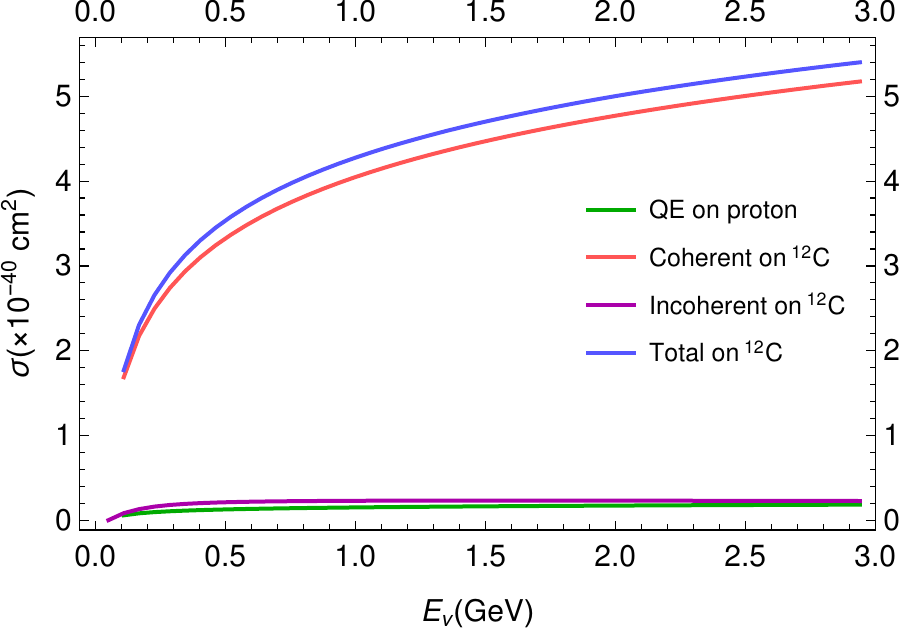}
        \includegraphics[width=0.48\textwidth]{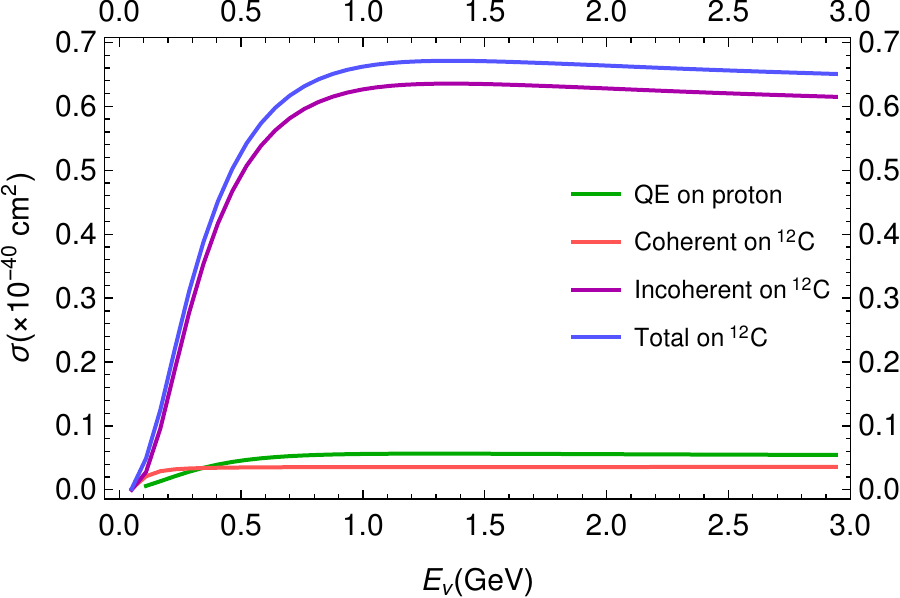}
    \caption{Integrated cross sections for $\nu_h$ production in $\nu_\mu$-nucleus scattering by EM (left) and weak (right) interactions as a function of the incident neutrino energy.}
    \label{fig:EM_NC_Cross_Sections}
\end{figure}
As shown in Fig.~\ref{fig:EM_NC_Cross_Sections}, with the set of parameters proposed in Ref.~\cite{Masip:2012ke}: $m_h = 50$~MeV, $\mu_{tr}^\mu=2.4\times 10^{-9} \mu_B$ and $|U_{\mu h}|^2 =0.003$, the EM cross section on nuclei is dominated by the coherent mechanism, while the incoherent one is suppressed by Pauli blocking at low four-momentum transfers, where the amplitude is enhanced by the photon propagator. On the contrary, the incoherent reaction is the largest contribution to the weak NC part. Interference terms between the EM and NC amplitudes are allowed but negligible.

The $\nu_h$ propagation and radiative decay inside the MiniBooNE detector has been investigated in Ref.~\cite{Edutesis}. We have taken advantage of the fact that, as pointed out in Ref.~\cite{Masip:2012ke}, the beam energies are large compared to $m_h$ and only an insignificant amount of the electromagnetically (weakly) produced heavy neutrinos have the spin against (aligned with) its momentum. From the effective Lagrangian of Eq.~\ref{eq:Masip_L} the angular distribution of the $\nu_h$ decay is found to be
\begin{equation}
    \frac{d \Gamma}{d \cos \theta_\gamma} (\nu_h \raw \nu_i + \gamma) = \frac{\left( \mu_\text{tr}^i \right)^2 m_h^3}{32 \pi} \left( 1 \mp \cos \theta_\gamma \right)
    \:,
    \label{eq:decay_width}
\end{equation}
where the negative (positive) sign corresponds to the decay of left (right)-handed heavy neutrino; relatives signs are reversed for antineutrinos. This result implies that photons from radiative decays are emitted predominantly in the direction opposite to the $\nu_h$ spin and along the direction of the $\bar\nu_h$ spin~\cite{Masip:2012ke}. In order to obtain the number of photon events in the detector and their angular and energy distributions one should further take into account that heavy neutrinos are produced with a scattering angle with respect to the incoming flux and travel a distance before their decay, which might occur outside the fiducial volume. Finally, to compare to the measured excess of events, the detection efficiency~\cite{webpage} has to be taken into account. 

The model is constrained by four parameters: the heavy neutrino mass, $m_h$; the mixing angle between light and heavy neutrinos, $U_{l h}$;
the heavy neutrino mean lifetime, $\tau_h$, which is related to the magnetic dipole moment through Eq.~\ref{eq:decay_width}, 
and the branching ratio of the $\nu_h$ decay to a light neutrino of flavor $i$, which depends on the corresponding transition  magnetic moments, $\mathrm{BR}_i = \left( \mu_\mathrm{tr}^i \right)^2/\sum_i \left( \mu_\text{tr}^i \right)^2$. 

With the parameters proposed in Ref.~\cite{Masip:2012ke}, the number of low energy events is underestimated in $\nu$-mode, while the agreement is good in $\bar\nu$-mode. However, the predominantly EM coherent contribution is strongly forward peaked, leading to a very narrow angular distribution not observed in the experiment (see Fig.~2 of Ref.~\cite{Alvarez-Ruso:2017hdm}). This result is in line with the findings of Ref.~\cite{Radionov:2013mca}.

\begin{figure}[h!]
  \centering
    \includegraphics[width=.47\textwidth]{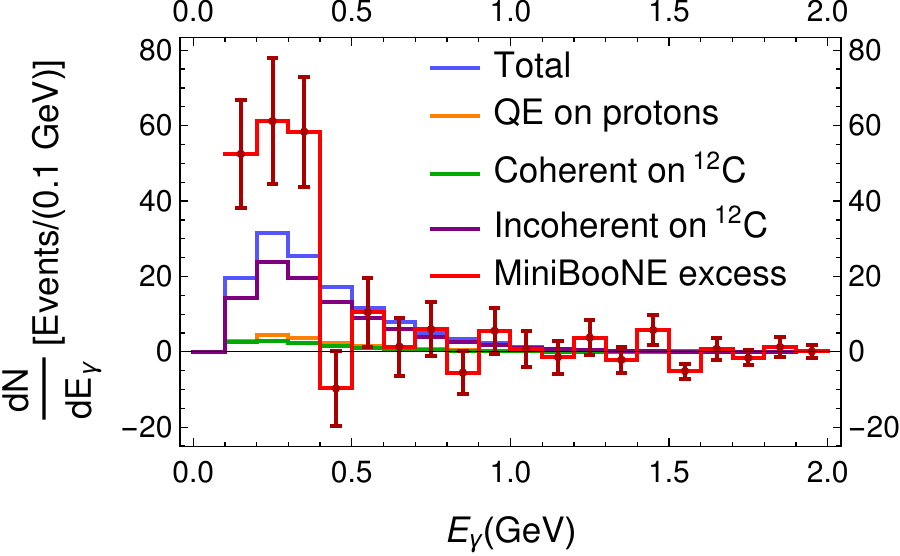}     
 $\qquad$
    \includegraphics[width=.45\textwidth]{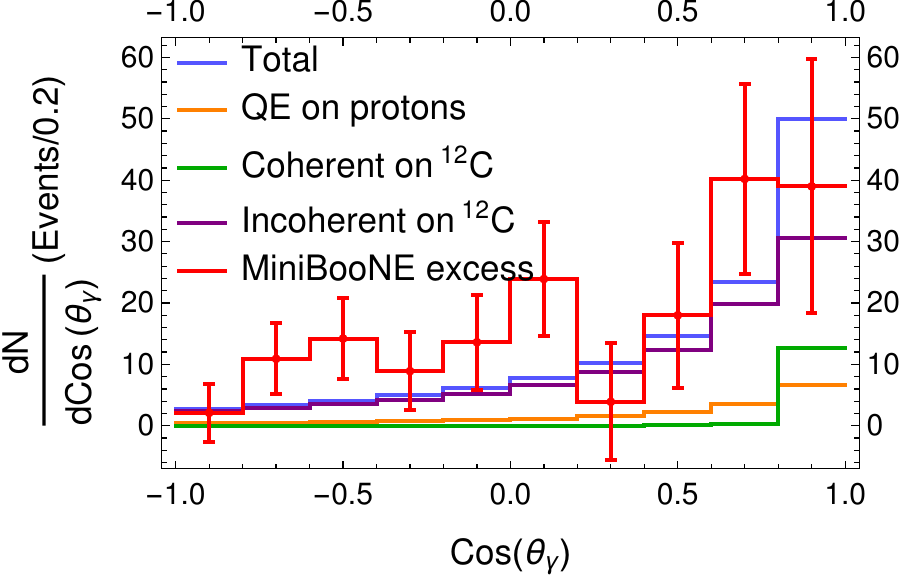}
  
    \includegraphics[width=.47\textwidth]{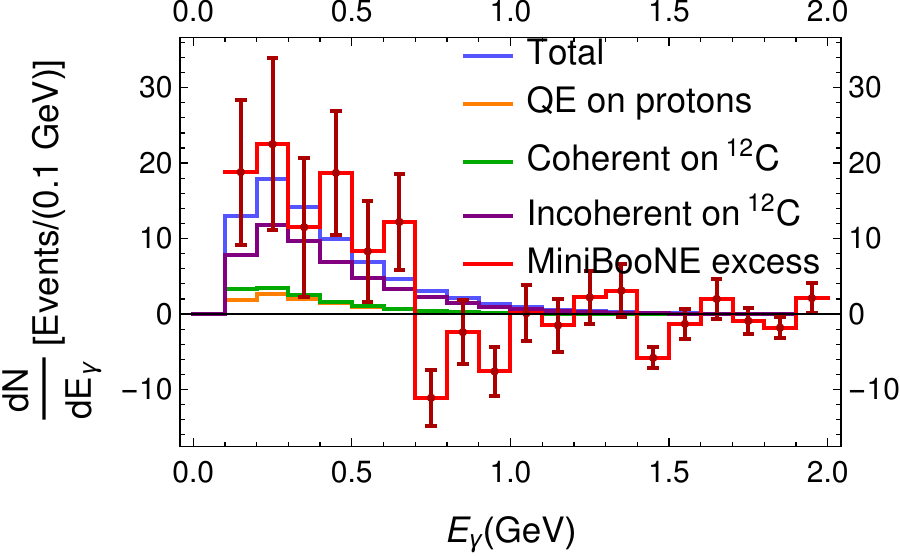}     
 $\qquad$
    \includegraphics[width=.45\textwidth]{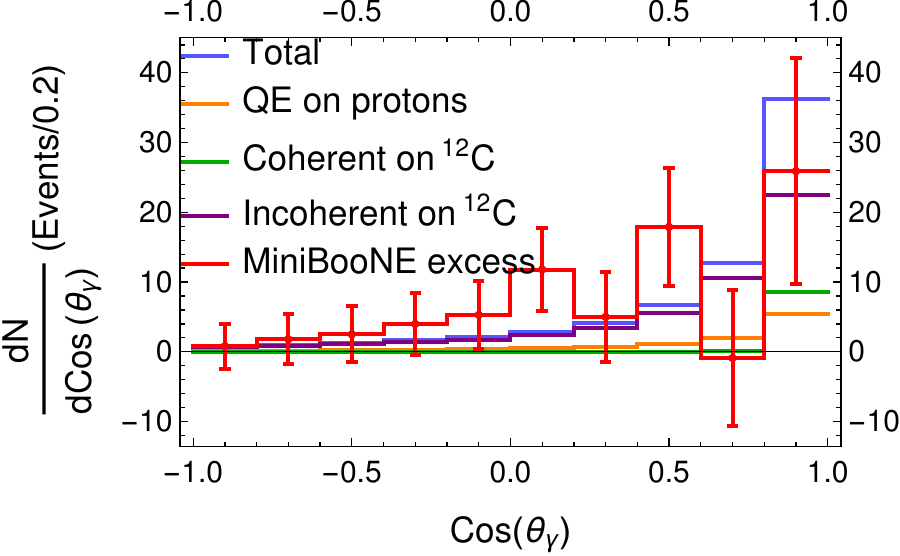}
  \caption{Photon events from radiative decay of $\nu_h$, $\bar\nu_h$ at the MiniBooNE detector in $\nu$-mode (top) and $\bar\nu$-mode (bottom) compared to the MiniBooNE excess. The individual contributions of different $\nu_h$ production mechanisms are shown.} 
  \label{fig:events2}
\end{figure}
The agreement can be improved by fitting the parameters in the allowed range established in Ref.~\cite{Gninenko:2010pr}. With values, $m_h =  70^{+ 10}_{- 30} $~MeV, BR$_\mu = 9^{+ 31}_{- 9} \times 10^{-4}$, $|U_{\mu h}|^2 =0.01$ and  $\tau_h = 2.5^{+ 0.6}_{- 1.2} \times 10^{-9}$ seconds, the  resulting event energy and angular distributions are given in Fig.~\ref{fig:events2}. The MiniBooNE excess of events is now better described, particularly the angular distributions. This is achieved at the price of reducing the EM strength, while increasing the NC one by setting $|U_{\mu h}|$ to its maximal allowed value: this upper limit in $|U_{\mu h}|$ prevents from obtaining a more satisfactory description of the data.   More stringent bounds for $U_{\mu h}$ exist, in particular from radiative muon capture: $\mu^- \, p \rightarrow n \, \nu \, \gamma$, experimentally investigated at TRIUMF. The mixing upper bound from Ref.~\cite{McKeen:2010rx} is a decreasing function of $m_h$ in the range under consideration (see Fig.~4 of  Ref.~\cite{McKeen:2010rx}). Once the fit results are largely independent on the mass (Figure 4.16 of Ref.~\cite{Edutesis}) and improve for larger values of $U_{\mu h}$, we can set the mass to its allowed minimum of 40 MeV in order to have the largest possible upper bound in the mixing: $|U_{\mu h}|^2 = 8.4 \times 10^{-3}$. The fit with these new restrictions finds $\tau_h = 9.1^{+ 1.1}_{- 1.5} \times 10^{-10}$ seconds, BR$_\mu = 1.7^{+ 2.4}_{- 1.4} \times 10^{-5}$ with a $\chi^2/$DoF only slightly above the one for the previous fit. 

The study outlined in this section shows that the hypothesis of Refs.~\cite{Gninenko:2009ks,Masip:2012ke} cannot satisfactorily explain the MiniBooNE anomaly. In particular, there are clear difficulties to simultaneously describe the energy and the angular distributions of the electron-like events. Nevertheless, based on MiniBooNE data, radiative decay of heavy neutrinos cannot be fully excluded at least as a partial source of the excess. It is worth studying it further in the new generation of experiments at the Booster Neutrino Beam, which should be able to distinguish photons from electrons. 

\paragraph{Production of sterile/dark neutrinos by beyond SM mediators.} The previous scenario in which a heavy (but relatively light) neutrino is produced in interactions with SM mediators has the advantage that the hadronic/nuclear part of the interaction is either well known or can be further constrained with neutrino scattering on nucleons and nuclei. On the other hand, as illustrated above, SM constrains restrict our ability to explain the anomaly in this way. Stimulated by the MiniBooNE updates confirming the event excess, several generalizations and extensions that avoid some of the bounds have been recently put forward. 

The basic idea of Ref.~\cite{Ballett:2018ynz} is that heavy ($100 \lesssim m_h \lesssim 250$~MeV) sterile neutrinos are produced at the detector by NC interactions mediated by a new GeV-scale boson and subsequently decay into an $e^+ e^-$ pair misreconstructed as an electron. The new boson, with a mass $m_{Z'} \gtrsim 1$~GeV is light enough to generate a large cross section with natural couplings and without the need of a large mixing $U_{\mu h}$, but heavy enough to elude the unrealistically narrow angular distribution that results in EM interactions. Interactions are built by extending the SM group with a $U(1)'$ gauge symmetry assumed to be broken at low energies. The low energy Lagrangian is~\cite{Ballett:2018ynz} 
\begin{align*} 
\mathcal{L} &= \mathcal{L}_{\nu\text{SM}}
-\frac{1}{4}X_{\mu\nu}X^{\mu\nu} - \frac{\sin\chi}{2}X_{\mu\nu}B^{\mu\nu} +
\frac{\mu^2}{2}X_\mu X^\mu \,;
\end{align*}
$\mathcal{L}_{\nu\text{SM}}$ denotes an extension of the SM incorporating
neutrino masses. The third term accounts for kinetic mixing characterized by the $\chi$ parameter; $F_{\mu\nu}\equiv\partial_\mu
F_\nu - \partial_\mu F_\nu$ where $F_\mu$ stands for $B_\mu$ the SM $U(1)_Y$ gauge field, and $X_\mu$ for the $U(1)'$ one with mass $\mu$ from symmetry breaking. The kinetic mixing term between $B_\mu$ and $X_\mu$ can be removed by
a field redefinition, identifying, after a change of basis, the states with definite mass with the photon, $Z$ and $Z'$ bosons. The coupling between SM fermions and the $Z^\prime$ is purely vector and proportional to both $\chi$ and the particle electric charge. SM-gauge 
singlets,  which are charged under U$(1)^\prime$, are introduced and mixed with the SM neutrinos. Therefore, one can have $Z'$-mediated $\nu_i \leftrightarrow \nu_h$ transitions 
\begin{equation}
    \mathcal{L}\supset U^*_{i h} g^\prime \, \overline{\nu_i}\gamma^\mu (1 - \gamma_5) \nu_h \, Z^\prime_\mu  \, + \, \mathrm{h.c.} \,,
\end{equation}
responsible for both the $\nu_h$ production in the scattering of incoming neutrinos off the target nuclei and its subsequent decay $\nu_h \raw \nu_i \, Z' \raw \nu_i \, e^+ \, e^-$. In Ref.~\cite{Ballett:2018ynz}, coherent scattering off $^{12}$C and incoherent scattering off constituent protons are considered but no medium corrections are applied to the later. 

\begin{figure}[h!]
    \centering
     \includegraphics[width=0.96\textwidth]{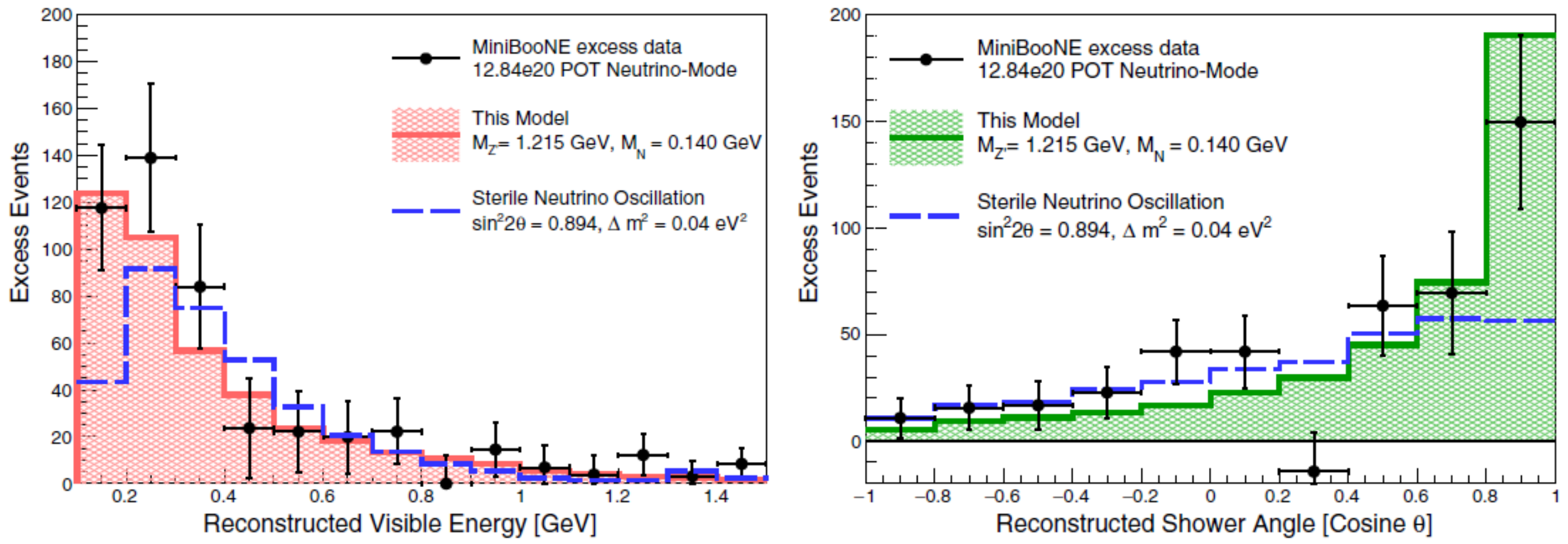}
    \caption{Model predictions of Ref.~\cite{Ballett:2018ynz} for the MiniBooNE event excess.}
    \label{fig:heavynuzp}
\end{figure}

The shape of the visible energy and angular distributions of the MiniBooNE excess are well described, as illustrated in Fig.~\ref{fig:heavynuzp} for typical $m_h = 140$~MeV and $m_{Z'} = 1.25$~GeV. 
The total number of events depends meanly on $\chi$ and the mixing angles. In Fig.~\ref{fig:heavynuzp} the authors of Ref.~\cite{Ballett:2018ynz} propose a minimal realization with $|U_{\mu h}|^2 = 1.5\times10^{-6}$, $|U_{\tau h}|^2 = 7.8\times10^{-4}$,  $\chi^2 = 5\times10^{-6}$ and $g^\prime=1$. Notice that in this scenario, a good agreement with data does not require large mixing angles as in the much more constrained case of SM mediators discussed previously. It should be nonetheless noted that the large $|U_{\tau h}|^2$ (520 times larger than  $|U_{\mu h}|^2$) would lead to sizable $\nu_h$ production rates in experiments with a large $\nu_\tau$component~\cite{Ballett:2019pyw}.

A successful explanation of the MiniBooNE anomaly is also obtained in Ref.~\cite{Bertuzzo:2018itn} in a model similar to the one just described but with the assumption of $m_{Z'} < m_h$ (instead of $m_{Z'} > m_h$ in Ref.~\cite{Ballett:2018ynz}) according to which the dark neutrino decays into an on-shell $Z'$. In this case, the benchmark parameters are $m_h = 420$~MeV and $m_{Z'} = 30$~MeV, with a significantly lighter $Z'$. However, according to the simulation of Ref.~\cite{Arguelles:2018mtc}, a model with such a light $Z'$ would lead to a much narrower angular distribution of the event excess than the one observed by MiniBooNE. This is in line with the previously discussed EM (photon exchange) scenario.       

These models, and variations including a larger number of heavy neutrinos~\cite{Abdullahi:2020nyr} or scalar mediators~\cite{Datta:2020auq, Abdallah:2020biq}, 
can be tested by the Short Baseline program (SBN) at Fermilab whose liquid argon detectors (SBND, MicroBooNE and Icarus) can distinguish between electrons and photon or $e^+ e^-$ showers. The relevant parameter space can also be probed at MINERvA and CHARM-II~\cite{Arguelles:2018mtc}. The authors of Ref.~\cite{Arguelles:2018mtc} stress that in the case of MINERvA, such a study would benefit from theoretical calculations of coherent $\pi^0$ and single-photon emission more suitable at higher energies [above the $\Delta(1232)$]. The model described in Sec.~\ref{sec:ncgamma} makes progress in this direction.  


\section{Outlook}
The anomalous excess of events found by MiniBooNE remains an open problem that has puzzled physicists for over a decade. It may be a manifestation of still unknown sterile neutrinos or even new forces of nature. Its solution, therefore, has potential  implications for the standing paradigm of neutrino and particle physics. On the other hand, an explanation related to unaccounted or poorly modeled backgrounds cannot yet be discarded, which calls for a better understanding of the interactions of few-GeV neutrinos with matter. New experimental information from the SBN program at Fermilab and the JSNS$^2$ at J-PARC but also from T2HK or DUNE will be priceless to close this chapter or, perhaps, open it wide.  

\paragraph{Acknowledgements.}
We are indebted to Matheus Hostert and Teppei Katori for their valuable comments about the manuscript. This research has been partially supported by the Spanish Ministerio de Ciencia e Innovaci\'on under contracts FIS2017-84038-C2-1-P and PID2020-112777GB-I00, the EU STRONG-2020 project under the program H2020-INFRAIA-2018-1, grant agreement no. 824093 and by Generalitat Valenciana under contract PROMETEO/2020/023.

\printbibliography

\end{document}